%% file: main.tex
\documentclass{aa}  

\usepackage{caption}
\usepackage{graphicx}
\usepackage{ulem}
\usepackage{txfonts}
\usepackage{color}
\definecolor{linkcolor}{rgb}{0,0,.7}
\definecolor{oldText}{rgb}{0.5,0,0}
\definecolor{newText}{rgb}{0,.5,0}

\usepackage[colorlinks=true]{hyperref}
\hypersetup{
  colorlinks   = true,
  urlcolor     = linkcolor,
  linkcolor    = linkcolor, 
  citecolor   = linkcolor 
}

\renewcommand{\u}{\underline}

\renewcommand{\doi}[1]{\textcolor{linkcolor}{\href{http://doi.org/#1}{\textbf{doi}}}}
\newcommand{\ads}[1]{\textcolor{linkcolor}{\href{#1}{\textbf{ads}}}}
\renewcommand{\arxiv}[1]{\textcolor{linkcolor}{\href{https://arxiv.org/abs/#1}{\textbf{arXiv}}}}

\newcommand{\e}[1]{\cdot 10^{#1}}

\newcommand{\unit}[2]{\,\text{#1}^{#2}}

\begin{document} 

   \title{The effect of nonlocal disk processes on the volatile CHNOS budgets of planetesimal-forming material}

   \author{M. Oosterloo
          \inst{1}
          \and
          I. Kamp
          \inst{1}
          \and
          W. van Westrenen
          \inst{2}
          }

   \institute{Kapteyn Astronomical Institute, University of Groningen, Landleven 12, 9747 AD Groningen, The Netherlands
                  \and
               Department of Earth Sciences, Vrije Universiteit Amsterdam, De Boelelaan 1085,
1081 HV Amsterdam, The Netherlands
               }

   \date{Received: 19 December 2023; accepted: 26 February 2024}

    \input{TextFiles/0Abstract}

    \keywords{}

    \maketitle

    \section{Introduction}
    \input{TextFiles/1Introduction.tex}
    \section{Methods}
    \input{TextFiles/2Methods.tex}

    \section{Results}
    \input{TextFiles/3Results.tex}

    \section{Discussion}
    \input{TextFiles/4Discussion.tex}

    \section{Conclusions and outlook}
    \input{TextFiles/5Conclusion.tex}
    
    \section*{Acknowledgements}
    This work is part of the second round of the Planetary and Exoplanetary Science Network (PEPSci-2), funded by the Netherlands Organization for Scientific Research (NWO). We would also like to thank an anonymous referee for his/her comments that helped to improve the manuscript.\\
    The numerical methods presented in this study were made possible through the use of Python packages Matplotlib \citep{Hunter2007}, NumPy \citep{vanderWalt+2011}, SciPy \citep{Virtanen+2020} and Pandas \citep{Pandas}.

%
%

\bibliographystyle{aa.bst}
\bibliography{refdata.bib}

\begin{appendix}
\input{TextFiles/6Appendix}
\end{appendix}

\end{document}

%% file: TextFiles/0Abstract.tex
 
  \abstract
   {The bulk abundances of CHNOS-bearing species of a planet have a profound effect on its interior structure and evolution. Therefore, it is key to investigate the behavior of the local abundances of these elements in the solid phase in the earliest stages of planet formation, where micrometer-sized dust grows into larger and larger aggregates. However, the physical and chemical processes occurring in planet-forming disks that shape these abundances are highly coupled and nonlocal.}
   {We aim to quantify the effects of the interplay between dynamical processes (turbulent diffusion, dust settling and radial drift), collision processes (coagulation and fragmentation), and the adsorption and desorption of ices on the abundances of CHNOS in local disk solids as a function of position throughout the planet-forming region.}
   {We used SHAMPOO (\u{S}toc\u{ha}stic \u{M}onomer \u{P}r\u{o}cess\u{o}r), which tracks the ice budgets of CHNOS-bearing molecules of a dust monomer as it undergoes nonlocal disk processing in a Class I disk. We used a large set of individual monomer evolutionary trajectories to make inferences about the properties of the local dust populations via a stochastic analysis of 64 000 monomers on a preexisting spatial grid.}
   {We find that spatially, monomers can travel larger distances farther out in the disk, leading to a larger spread in positions of origin for a dust population at, for example, $r=50$ AU compared to $r=2$ AU. However, chemically, the inner disk ($r\lesssim 10$ AU) is more nonlocal due to the closer spacing of ice lines in this disk region. Although to zeroth order the bulk ice mantle composition of icy dust grains remains similar compared to a fully local dust population, the ice mass associated with individual chemical species can change significantly. The largest differences with local dust populations were found near ice lines where the collisional timescale is comparable to the adsorption and desorption timescales. Here, aggregates may become significantly depleted in ice as a consequence of microscopic collisional mixing, a previously unknown effect where monomers are stored away in aggregate interiors through rapid cycles of coagulation and fragmentation.}
   {Nonlocal ice processing in a diffusion-dominated, massive, smooth disk has the most significant impact on the inner disk ($r\lesssim 10$ AU). Furthermore, microscopic collisional mixing can have a significant effect on the amounts of ice of individual species immediately behind their respective ice lines. This suggests that ice processing is highly coupled to collisional processing in this disk region, which implies that the interiors of dust aggregates must be considered and not just their surfaces.}

%% file: TextFiles/1Introduction.tex
The chemical elements carbon (C), hydrogen (H), nitrogen (N), oxygen (O), and sulfur (S) are important elements for the chemical habitability of rocky planets \citep[][]{Krijt+2022}. For example, they are the fundamental building blocks of life itself, being the main constituents of many biologically relevant molecules \citep[][]{Baross+2020, Sasselov+2020}. CHNOS-bearing molecules can also affect the physical habitability of planets \citep[e.g. ][]{Krijt+2022}. For example, the amount of CHNOS bearing molecules (such as CO$_2$, H$_2$O, and N$_2$) present in the planetary atmosphere plays a key role in determining the width of the circumstellar habitable zone where liquid water can exist on the planetary surface \citep[][]{Kasting+1993, Kasting&Catling2003, Kopparapu+2013}. The structure and evolution of the planetary interior are also strongly affected by the planetary budgets of CHNOS. Here, CHNOS abundances have important effects on the size and structure of the planetary core \citep[e.g.,][]{Tronnes+2019, Johansen+2023}, the physical structure and mineralogy of their mantles \citep[e.g.,][]{Kushiro+1969, Dasgupta&Hirschmann2006, Hakim+2019}, and on the chemical composition of volcanic outgassing \citep{Bower+2022}, which in turn has significant effects on the long-term evolution of the atmospheric composition \citep[e.g.,][]{Tosi+2017, Oosterloo+2021}.\\
Altogether, it is thus key to identify how much CHNOS a nascent planet receives from its host planet-forming disk. At the onset of planet formation, most solid material exists as millimeter- to centimeter-sized dust. These dust grains are subject to transport processes as a consequence of their interaction with the surrounding gas \citep[e.g.,][]{Armitage2010}. For small grains (i.e. grains with a Stokes number St$\ll 1$) transport primarily occurs through turbulent diffusion, whereas the dynamical behavior of grains with a large Stokes number is dominated by aerodynamic drag, resulting in dust settling and radial drift \citep[e.g.,][]{Weidenschilling1977, Armitage2010}. This dynamical transport has been shown to allow individual dust grains to be exposed to a wide range of local physical conditions in protoplanetary disks \citep[][]{Ciesla2010, Ciesla2011}{}. \\ 
In colder disk regions, a considerable fraction of the solid-phase CHNOS mass budget in this first stages of planet formation may be incorporated as ices, with H$_2$O, CO, CO2, CH$_4$, NH$_3$, H$_2$S, OCS, and SO$_2$ being major carrier molecules \citep[][]{Boogert+2015, Krijt+2020,Oberg&Bergin2020}. In addition, for carbon and oxygen, up to 50\% of the total elemental mass budget can be locked away in more refractory solid material such as amorphous carbon (graphite) and silicates, respectively \citep[][]{Mishra&Li2015, Oberg&Bergin2020}. Large amounts of nitrogen are likely also stored into the hypervolatile, chemically inert N$_2$ at the onset of planet formation, which may contribute to the nitrogen deficiency in comets \citep[e.g.,][]{Bergin+2015, Furuya+2018, Cleeves+2018, Altwegg+2020}. The main reservoirs for sulfur during the first stage of planet formation are less well understood. Although a large fraction of S in planet-forming disks may also exist in a solid-phase reservoir, the precise nature of this reservoir remains unknown, with currently detected ices amounting to $\sim 5\%$ of the total sulfur elemental budget at most \citep[][]{Boogert+2015, Kama+2019, LeGal+2021, Keyte+2024}. \\
Although a considerable fraction of the solid-phase CHNOS budget exists as ices on dust grain surfaces, the processes that determine the amounts of ice present on dust grains, adsorption and desorption, vary strongly as a function of the local temperature, radiation field, and gas phase composition \citep[see e.g.][for a review]{Cuppen+2017}. Since dynamical transport processes can expose individual dust grains to a wide range of physical conditions, it is key to assess the systematic effects of these dynamical processes on the volatile\footnote{\textit{Volatile} CHNOS signifies the part of all the solid-phase CHNOS that is incorporated in ices.} CHNOS budgets of local dust populations. However, disentangling dynamical processes from collisional and ice processes is problematic as the disk regions where the timescales associated with these processes are similar depend on grain size and molecular species \citep[][]{Oosterloo+2023}{}. The volatile CHNOS budget is usually determined by a set of molecules, while different grain sizes are connected via collisional processes. This issue has previously motivated the development of the SHAMPOO\footnote{\url{https://github.com/moosterloo96/shampoo}} code (\u{S}toc\u{ha}stic \u{M}onomer \u{P}r\u{o}cess\u{o}r) in \cite{Oosterloo+2023}, which tracks the effects of dynamical, collisional and ice processing on the volatile CHNOS budgets of a single tracer particle called a "monomer" traveling throughout a pre-calculated disk environment calculated with the thermochemical disk model ProDiMo \citep{Woitke+2009, Kamp+2010, Thi+2011, Thi+2013}.\\
In this work, we aim to quantitatively constrain the nonlocal effects originating from vertical settling, radial drift and turbulent diffusion on the volatile CHNOS budgets of local dust at the onset of planetesimal formation. We use trajectories of 64 000 monomers generated with the SHAMPOO code to quantitatively investigate nonlocal effects on the volatile CHNOS budgets of local dust populations. \\
This paper is organized as follows: In Sect. \ref{sec:2} we present a short overview of ProDiMo and SHAMPOO, and our approach to inferring local dust properties from a large set of monomer simulations. Subsequently, we present our results in Sect. \ref{sec:3}, and note their implications for planet formation and compare them to earlier work in Sect. \ref{sec:4}, while we summarize our main conclusions in Sect. \ref{sec:5}.

%% file: TextFiles/2Methods.tex
\label{sec:2}

\subsection{Background disk model}
\label{sec:2.1}
\input{TextFiles/Section2/Section21}

\subsection{The SHAMPOO code}
\label{sec:2.2}
\input{TextFiles/Section2/Section22}

\subsection{From monomers to dust populations}
\label{sec:2.3}
\input{TextFiles/Section2/Section23}

%% file: TextFiles/Section2/Section21.tex
\input{Tables/21DiskProperties.tex}
\input{Tables/21AdsorptionEnergies.tex}
In SHAMPOO, individual monomers are allowed to travel throughout a static disk environment. As in \cite{Oosterloo+2023}, we utilize the thermochemical disk model ProDiMo to specify the physical and chemical structure of a background protoplanetary disk on a prespecified grid of radial and vertical positions $r_i,z_j$. Linear interpolation of these quantities subsequently allows for the estimate of physical and chemical condition at any given monomer position $r,z$.\\
For the physical disk structure, SHAMPOO requires quantification at positions $r_i,z_j$ of the gas density $\rho_\text{g}$, dust density $\rho_\text{d}$, gas temperature $T_\text{g}$, dust temperature $T_\text{d}$ and the local UV radiation field $\chi_\text{RT}$ as a function of radial and vertical position throughout the entire disk, where we assume the disk is axisymmetric around the vertical $z$-axis. $\rho_\text{g}$ and $\rho_\text{d}$ are calculated at positions $r_i,z_j$ from the parametrized density structure presented in \cite{Woitke+2016} using the normalization methodology outlined in \cite{Woitke+2016} and \cite{Oosterloo+2023}. $T_\text{g}$, $T_\text{d}$ and $\chi_\text{RT}$ are determined from the local radiation field $J_\nu$, which is calculated in ProDiMo by solving the local continuum radiative transfer. For a full description of the treatment of radiative transfer and the heating and cooling processes in ProDiMo we refer the interested reader to \cite{Woitke+2009, Woitke+2016, Thi+2011, Aresu+2011} and \cite{Oberg+2022}.\\
For the ice evolution of monomers we also require the number densities $n_x$ associated with molecule species $x$ at positions $r_i,z_j$ for the adsorption rates of various molecules. For this purpose we perform time-dependent chemistry in the same two-step approach as in \cite{Oosterloo+2023}. The time-dependent chemistry is solved for the large DIANA chemical network, which contains 13 elements and 235 species \citep{Kamp+2017}. We take into account gas-phase chemistry and the adsorption and desorption of volatile ices, but do not account for surface chemistry. This may result in lower estimates for the amounts of species that primarily form through hydrogenation, in particular CH$_3$OH \citep[e.g.,][]{Cuppen+2009, Yu+2016, Bosman+2018b}. For the adsorption energies, we use the values from the 2012 edition of the UMIST database \citep{McElroy+2013}, which are also listed for the key species in this study in Table \ref{tab:AdsorptionEnergies}. For the discussion of this specific chemical network, we refer the reader to \cite{Kamp+2017}, while the treatment of chemistry in ProDiMo is discussed in \cite{Woitke+2009, Aresu+2011}.\\
Throughout this work, we use the model parameters associated with the \texttt{vFrag1}-model from \cite{Oosterloo+2023}. A summary of the key physical properties of the disk in this model is presented in Table \ref{tab:DiskProperties}, while we refer the interested reader to \cite{Oosterloo+2023} for a full table of model parameters and an overview of the calculated physical and chemical structures of this disk model. In the \texttt{vFrag1}-model, the maximum grain size $a_\text{max}(r)$ is consistent with a fragmentation velocity $v_\text{frag}=1\,\unit{m}{}\unit{s}{-1}$. The disk parameters are chosen to reflect a young class I disk. However, we do not include an envelope structure typical for class I sources \citep[e.g.,][]{Armitage2015, Fischer+2023}{}{}. Instead, $\rho_\text{g}$ and $\rho_\text{d}$ taper off exponentially in the background model for $r>100$ AU (Table \ref{tab:DiskProperties}). Altogether, the background disk model may be inappropriate to model the evolution of dust populations exterior to 100 AU in this evolutionary stage. We therefore restrict our analysis in this work to radial positions $r<100$ AU, while we discuss the differences between class I and class II disks and their possible effects on the results in this work in Sect. \ref{sec:4.3}.

%% file: Tables/21DiskProperties.tex
\begin{table}[]
\begin{center}
    \begin{tabular}{clrl}
    \hline
         \textbf{Parameter}& \textbf{Name} &\textbf{Value} & \textbf{Unit} \\\hline\hline
$M_\text{disk}$ & Disk mass                     & 0.1 & $M_\odot$                            \\
$\delta$    & Dust to gas mass ratio            & 0.01          & -         \\
$r_\text{in}$ & Inner disk radius               & 0.07          & AU        \\
$r_\text{out}$ & Outer disk radius              & 600           & AU                         \\
$r_\text{taper}$ & Tapering-off radius          & 100           & AU \\
$\epsilon$ & Column density exponent            & 1             & -       \\
$\alpha$ & Turbulence strength parameter        & $10^{-3}$     & -                          \\
$M_\star$ & Stellar mass                        & $0.7$     & $M_\odot$     \\
$L_\star$ & Total stellar luminosity            & $6$       & $L_\odot$     \\
$T_\text{eff}$ & Stellar effective temperature  & $4000$    & K              \\
\hline
\end{tabular}
\caption{Summary of the key disk properties of the \texttt{vFrag1} background disk model in this work. For the full list of model parameters, we refer the interested reader to \cite{Woitke+2016} and \cite{Oosterloo+2023}.}
\label{tab:DiskProperties}
\end{center}
\end{table}

%% file: Tables/21AdsorptionEnergies.tex
\begin{table}[]
\begin{center}
\begin{tabular}{cccc}
Species $x$ & $E_{\text{ads},x}/k_\text{B}$ (K) & Species $x$ & $E_{\text{ads},x}/k_\text{B}$ (K) \\\hline\hline
H$_2$O      & 4800  &   NH$_3$      & 5534     \\
CO          & 1150  &   H$_2$S      & 2743     \\                           
CO$_2$      & 2990  &   SO$_2$      & 5330     \\
CH$_4$      & 1090  &   OCS         & 2888     \\
CH$_3$OH & 4930 & & \\\hline

\end{tabular}
\caption{Adsorption energies used for the molecule species considered throughout this work \citep{McElroy+2013}.}
\label{tab:AdsorptionEnergies}
\end{center}
\end{table}

%% file: TextFiles/Section2/Section22.tex
The SHAMPOO code is a stochastic model which tracks the effects of dynamical transport, collisional processes and ice processing on the volatile CHNOS budget of an individual tracer particle called a monomer. A monomer effectively represents a spherical unit of mass of fixed radius $s_\text{m}=5\e{-8}$ m. Due to collisional growth, the monomer is usually embedded at a depth $z_\text{m}$ inside a much larger aggregate whose size $s_\text{a}$ changes over time due to coagulation and fragmentation. The aggregate which contains the monomer tracked by SHAMPOO at given time $t$ is referred to as the home aggregate. The dynamical behavior of the monomer depends on the home aggregate size $s_\text{a}$, whereas the monomer depth $z_\text{m}$ influences the efficiency of adsorption and photodesorption of molecules on the monomer surface. Both $s_\text{a}$ and $z_\text{m}$ evolve over time due to the collisions between the home aggregates and other dust particles, with collision rates being informed from the background disk model. For the ice evolution of individual monomers SHAMPOO evaluates a set of ice mass balance equations during every timestep $\Delta t$ to track the gain and loss of different ice species from the monomer ice mantle. The ice mantle of an individual monomer is here assumed to be well-mixed. Monomers close to the surface of their home aggregate at low depth $z_\text{m}$ are allowed to undergo adsorption, thermal desorption and photodesorption, whereas at depths greater than a critical monomer depth $z_\text{crit}= 2s_\text{m}=10^{-7}\,\unit{m}{}$, monomers are assumed to be shielded by other monomers in the home agggregate from gas phase molecules and UV photons, which means that monomers located at large $z_\text{m}$ are only allowed to undergo thermal desorption \citep{Oosterloo+2023}. We note that in aggregates with high filling factors, thermally desorbed molecules can take significant time to escape the aggregate interior, as collisions with other monomers may result in the re-adsorption of these molecules by other monomers. We discuss the potential effects of readsorption and molecule diffusion in the deep aggregate interior can have on the results presented of this work in Sect. \ref{sec:4.3}. For a more elaborate description of the SHAMPOO code and its treatment of dynamical, collisional and ice processing we refer the interested reader to \cite{Oosterloo+2023}. Furthermore, we discuss improvements in the SHAMPOO code with respect to the version discussed in \cite{Oosterloo+2023} in Appendix \ref{sec:AB}. 

%% file: TextFiles/Section2/Section23.tex
\begin{figure}[]
    \centering
    \includegraphics[width=.5\textwidth]{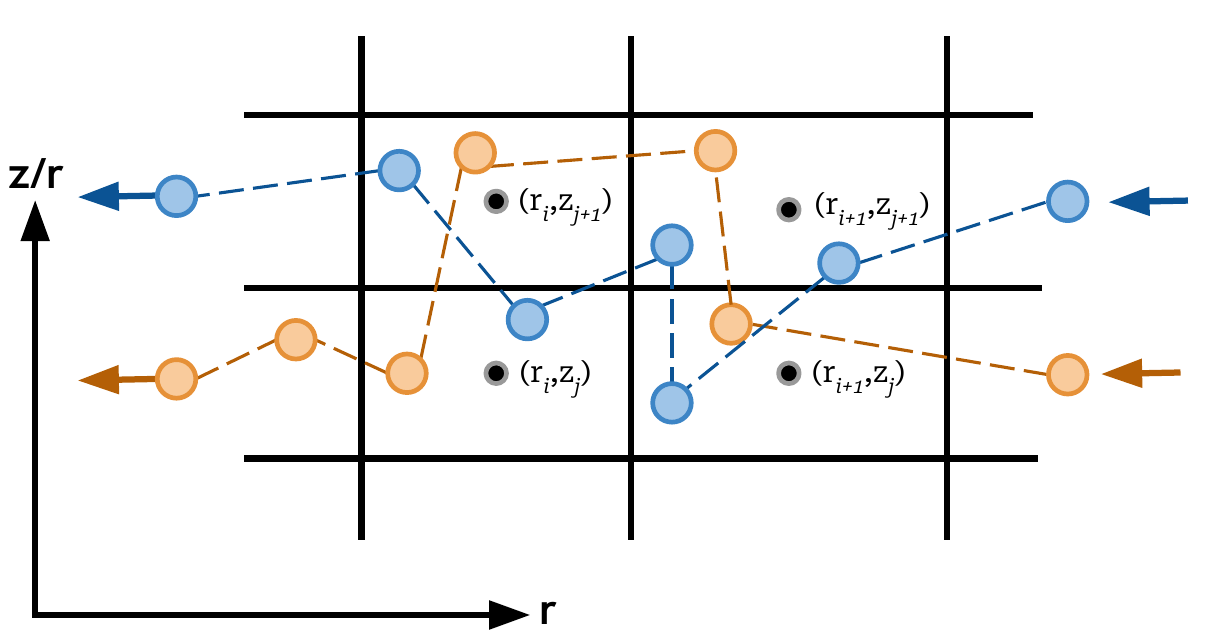}
    \caption{Illustration of the discretization of two example monomer trajectories on the background grid informed from background disk model coordinates $(r_i,z_j)$.}
    \label{fig:23MonomerStatisticsExample}
\end{figure}

In this section we outline our approach of converting the many monomer trajectories produced by SHAMPOO into statistics that represent the properties of local dust, such as the local amounts of ices.\\
Throughout this work, we consider the evolution of a single large set of 64000 monomers over 100 kyr. Individual monomers are treated separately from each other, such that the computational evaluation of individual monomers can be performed both parallel and sequentially. At each timestep in a monomer simulation, we track and update the monomer position $(r,z)$, home aggregate size $s_\text{a}$, monomer depth $z_\text{m}$, and the amount of ice of species $x$, denoted by $m_x$. Although each monomer is simulated for 100 kyr, the number of timesteps associated with a single monomer can vary substantially depending on dynamical and collisional timescales, and the cumulative number of timesteps for all monomers is $\sim 3.5\e{8}$.\\
In this set of simulations, we track the adsorption and desorption of H$_2$O, CO, CO$_2$, CH$_4$, CH$_3$OH, NH$_3$, H$_2$S, SO$_2$ and OCS on and from monomers, respectively. The associated number densities of these molecules in the gas and ice phase of the background disk model are shown in Fig. \ref{fig:AABackgroundModelAbundances}. H$_2$O, CO, CO$_2$, CH$_4$, CH$_3$OH and NH$_3$ are thought to be the main carrier molecules of ice-phase carbon, nitrogen and oxygen in disks, and have also been inferred to be the most abundant carrier molecules of these elements in cometary ices \citep[e.g.,][]{Bockelee-Morvan+2017, Oberg&Bergin2020}. However, the main volatile carrier molecules for sulfur in protoplanetary disks are poorly constrained. In protoplanetary disks, a number of gaseous sulfur species have been detected, most notably CS, SO and H$_2$S \citep[e.g.,][]{Booth+2018, Phuong+2018, LeGal+2019}. In the interstellar medium, OCS has been detected as ice, while detections of SO$_2$ ice have been only tentative so far \citep{Boogert+2015, Boogert+2022, McClure+2023}. Furthermore, modeling work suggests that HS-ice and H$_2$S-ice may be an important sulfur reservoir in dark clouds \citep{Vidal+2017}. Hence, for the carrier molecules of volatile sulfur, we consider the molecules OCS, SO$_2$ and H$_2$S, which are major volatile sulfur carrier molecules in comets \citep{Bockelee-Morvan+2017}.\\
For the monomer and home aggregate, we mostly use the same set of parameters as shown in Table 3 of \cite{Oosterloo+2023}. However, we here set the number density of adsorption sites on the monomer surface $N_\text{ads}=1.5\e{19}\,\unit{m}{-2}$ and the number of active monolayers $N_\text{act}=2$ in order to be consistent with the chemistry parameters in the background disk model \citep{Woitke+2009}. The change in $N_\text{ads}$ and $N_\text{act}$ does not significantly alter results with respect to the parameter choice used in \cite{Oosterloo+2023}, $N_\text{ads}=1\e{19}\,\unit{m}{-2}, N_\text{act}=3$. Furthermore, in this work, home aggregates are assumed to have a filling factor $\phi=1$, in order for the density properties of home aggregates to be consistent with dust properties in the background disk model.\\ 
The initial monomer positions are sampled from a loguniform and uniform distribution for $r$ and $z/r$, respectively. We define the boundaries of the sampling region as \hbox{$r\in [0.5,100]$ AU} and $z/r\in [-0.1,0.1]$. The initial home aggregate size $s_\text{a}$ is determined from the mass weighted, local dust size distribution, and the initial monomer depth $z_\text{m}$ is drawn from a spherically uniform distribution $z_m\in [0,s_\text{a}]$. The initial amounts of ice on the monomers are set equal to the local ice to dust ratios in the background disk model.\\
In the large nonlocal set of monomers, position coordinates associated with individual monomers can take any continuous value of $r$ and $z$ at each timestep, undergoing a random-walk motion throughout the disk according to the prescription provided in \cite{Ciesla2010, Ciesla2011}. In order to make inferences about local dust populations, we assign the position coordinate associated with every monomer timestep to the nearest spatial grid point in the background model. Effectively this background model grid provides a method of "binning" monomer positions into spatial cells surrounding each background model grid point, as illustrated in Fig. \ref{fig:23MonomerStatisticsExample}. The large number of monomers evaluated allows to gradually collect a data set consisting of many timesteps from different monomers in each spatial grid cell. These timesteps can be used as samples that are representative for the physical and chemical properties of the local dust. Local dust properties, such as the average amount of ice, mean aggregate size and mean monomer depth can be inferred by averaging over a large amount of samples. \\
Although binning the spatial coordinates associated with each monomer timestep allows for the construction of local data sets, there are a number of biases we need to account for via weighting when inferring average behavior from the data obtained from discretized monomers.
\\
The first applied weight associated with data point $n$ contributed by monomer $m$ is $w_{\Delta t,nm} = \Delta t_{nm}$, the timestep size, since monomers that reside longer in a certain spatial grid cell contribute more to the average properties of the dust in the respective grid cell than a monomer that resides in the same grid cell for a relatively short period. The second weight we apply is related to our bias in monomer sampling. Since monomers represent a unit of dust mass, the physical distribution of monomers is given by the dust mass distribution $\rho_\text{d}(r,z)$. However, the probability distribution for our random sampling does not follow $\rho_\text{d}$. This is in particular true in the $z$-direction as monomers are sampled uniformly up to $|z/r|=0.1$, while $\rho_\text{d}$ decreases strongly as a function of $z$. This difference between the physical distribution of monomers $P_\text{p}(r,z)\propto \rho_\text{d}(r,z)$ and the numerical distribution of monomers $P_\text{n}(r,z)$ obtained from our radial loguniform and vertical uniform sampling has to be accounted for. We therefore define the initial position probability correction weight $w_{\text{p},nm}=P_\text{p}(r_{0m},z_{0m})/P_\text{n}(r_{0m},z_{0m})$, where $r_{0m}$ and $z_{0m}$ denote the initial position of monomer $m$. Furthermore, in order for the distribution of monomers to be proportional to the physical mass distribution $\rho_\text{d}$, it was found that the application of an additional weight $w_{\text{g},nm}\propto 1/(\Delta r_{0m} \Delta z_{0m})$ was necessary. $\Delta r_{0m}$ and $\Delta z_{0m}$ here denote the width (in AU) and height (in units of $z/r$) of the grid cell of origin of monomer $m$.\\
Altogether we can write the expectation value $\langle X\rangle$ for monomer property $X$ in the spatial grid cell associated with $(r_i,z_j)$ as 
\begin{align}
    \label{eq:weighting}\langle X\rangle=\frac{\sum\limits_{m=1}^{N_\text{mon}}\sum\limits_{n=1}^{N_m}X_{nm}\cdot w_{\Delta t,nm}w_{\text{p},nm}w_{\text{g},nm}}{\sum\limits_{m=1}^{N_\text{mon}}\sum\limits_{n=1}^{N_m}w_{\Delta t,nm}w_{\text{p},nm}w_{\text{g},nm}}.
\end{align}
Here, $N_\text{mon}$ denotes the total number of monomers that contribute data points associated to a given grid cell, $N_m$ is the total number of timesteps monomer $m$ contributes to the grid cell, and $X_{nm}$ is the value of monomer property $X$ for monomer $m$ at timestep $n$.\\
In addition to the required weighting schemes highlighted above, testing revealed that near the inner boundary of the sampling region between $r=0.5$ AU and $r=1$ AU, a large number of monomers are lost to $r<0.5$ AU. These monomers are not replenished by monomers originating from $r<0.5$ AU due to the choice of sampling region, resulting in an artificial gradual loss of monomers from this disk region. Since the monomer population is thus incomplete in this disk region, we restrict our analysis of expected values to the region between $r=1$ AU and $r=100$ AU throughout this work. At the outer edge of the sampling region, the effects of this issue were found to be limited.\\
We note that the interpolated physical and chemical conditions (i.e. $\rho_\text{g}$, $n_x$, etc.) at the monomer position $(r,z)$ do not have to be equal to the physical conditions in the bin associated with grid point $(r_i,z_j)$. This means that the monomers used as representative particles at grid point $(r_i,z_j)$ in reality represent a small range of physical and chemical conditions throughout the grid cell associated with grid point $(r_i,z_j)$. However, with the grid resolution used in the background model (150 radial and 100 vertical grid points), relative variations are on the order of a few percent on average (Appendix \ref{sec:AC}). Furthermore, the 64 000 monomer trajectories considered in this work were found to result in a sufficient number of monomer data points in each spatial grid cell for $r\gtrsim1$ AU, $|z/r|\lesssim 0.1$ (Appendix \ref{sec:AD}).

%% file: TextFiles/3Results.tex
\label{sec:3}

\subsection{The nonlocal origin of local dust}
\label{sec:3.1}
\input{TextFiles/Section3/Section31}

\subsection{Exposure of monomers to the environment}
\label{sec:3.2}
\input{TextFiles/Section3/Section32}

\subsection{Ice composition: Local vs. nonlocal}
\label{sec:3.3}
\input{TextFiles/Section3/Section33}

\subsection{Behavior near ice lines}
\label{sec:3.4}
\input{TextFiles/Section3/Section34}

\subsection{Elemental ratios}
\label{sec:3.5}
\input{TextFiles/Section3/Section35}

%% file: TextFiles/Section3/Section31.tex
Throughout the entire sampling region, the dominant process facilitating transport of dust is turbulence-induced diffusion. This is mainly a consequence of the fact that we consider a relatively massive disk, which results in a lower Stokes number for a dust aggregate of given size \citep{Armitage2010}. We explore the Stokes number associated with the expected aggregate size $\langle s_\text{a}\rangle$ as a function of position throughout the sampling region in Appendix \ref{sec:AE} and find that throughout the majority of the sampling region, St$\lesssim 10^{-3}$ for the average dust aggregate.\\
\begin{figure}[]
    \centering
    \includegraphics[width=.45\textwidth]{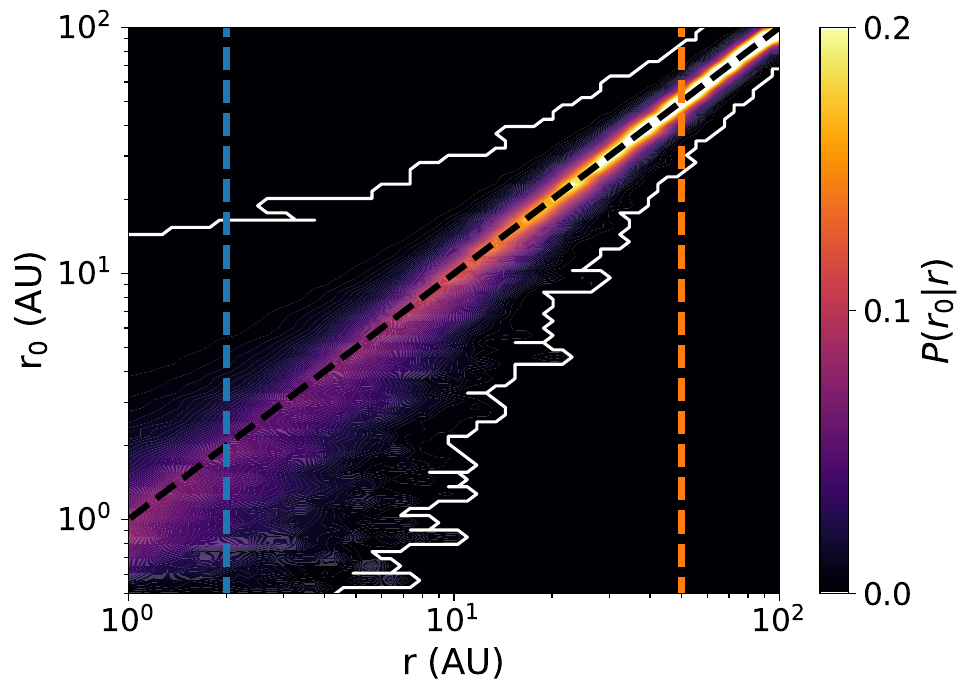}
    \caption{Distribution $P(r_0|r)$ of monomers originating from positions $r_0$ given radial position $r$. The dashed lines indicate the positions of the slices at $r=2,50$ AU shown in Fig. \ref{fig:31NonLocalF3_cleaned3OriginHists}.}
    \label{fig:31NonLocalF3_cleaned3OriginDiagram}
\end{figure}
\begin{figure}[]
    \centering
    \includegraphics[width=.45\textwidth]{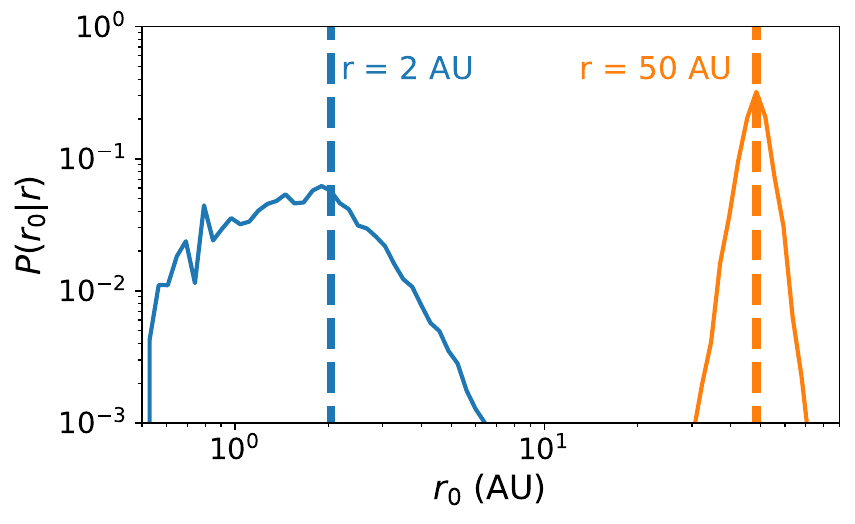}
    \caption{Slices of the monomer distribution $P(r_0|r)$ from Fig. \ref{fig:31NonLocalF3_cleaned3OriginDiagram} as a function of monomer positions of origin $r_0$ at $r=2$ AU and $r=50$ AU.}
    \label{fig:31NonLocalF3_cleaned3OriginHists}
\end{figure}
\begin{figure}[]
    \centering
    \includegraphics[width=.49\textwidth]{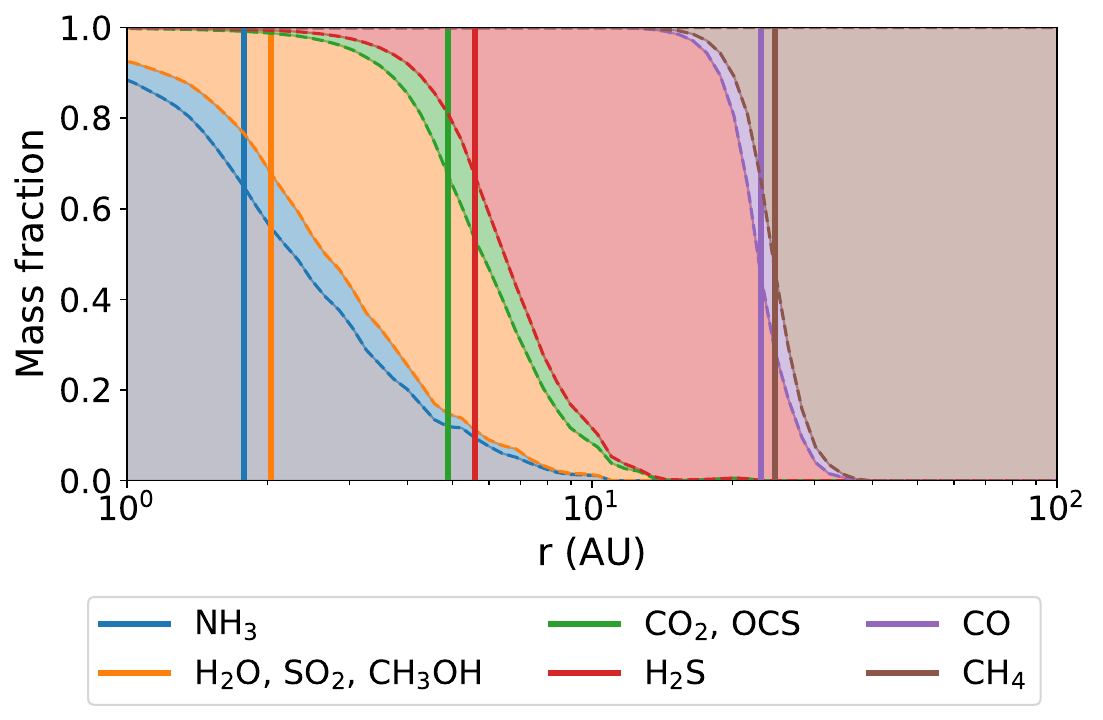}
    \caption{Weighted fraction of monomers at radial distance $r$ originating from regions behind different ice lines. The colored regions denote monomers originating from behind the respective ice line, whereas the gray zone in the lower left denotes monomers originating from interior to the NH$_3$ ice line.}
    \label{fig:31NonLocalF3_cleaned3ChemicalOrigin}
\end{figure}
In order to quantify the amount of nonlocal dust in local dust populations, we consider the radial position of origin $r_0$ associated with each monomer at every radial position between $r=1$ AU and $r=100$ AU. This gives rise the series of histograms shown in Fig. \ref{fig:31NonLocalF3_cleaned3OriginDiagram}. We here compare the data points of all grid cells at a given radial position $r$ with the radial position associated with the first data points of the monomers at $t=0$. It becomes clear that there is a nonlocal contribution to local dust populations at almost all radial distances. In Fig. \ref{fig:31NonLocalF3_cleaned3OriginDiagram} populations at $r\lesssim
10$ AU appear more nonlocal than at larger radii, but this is a projection effect from the logarithmic axes of Fig. \ref{fig:31NonLocalF3_cleaned3OriginDiagram}. This becomes clear when comparing the absolute spread in $r_0$ at $r=50$ AU with the spread at $r=2$ AU (Fig. \ref{fig:31NonLocalF3_cleaned3OriginHists}). Monomers at 2 AU are a mixture of monomers originating from $r<1$ AU to $r\approx 6.5$ AU, while at $50$ AU, monomers tend to come from 40-60 AU, but may also originate from as far in as 35 AU and as far out as 70 AU. Therefore, in terms of physical traversed distance, we find that dust is more nonlocal at larger $r$.\\
Chemically, transport processes have a stronger effect on the dust ice composition at $r\lesssim 10$ AU, since the ice lines of most species are located between 1 and 10 AU in our model. This becomes clear upon categorization of monomers based on their ice region of origin in Fig. \ref{fig:31NonLocalF3_cleaned3ChemicalOrigin}. We here categorize monomers in grid cells at the same radial distance based on their position of origin relative to the ice lines in the disk model.\footnote{The ice line of a species is here defined to be located at the position in the midplane where more than $50\%$ of the total amount of molecules of a chemical species exists as ice in the background disk model.} Monomers at smaller $r$ originate from chemically more diverse environments than at larger radii, mainly as a consequence of the closer spatial distance between the ice lines of individual chemical species. For example, at $r=3$ AU, more than 30 \% of the monomers that have visited $r=3$ AU over the 100 kyr time period originated from inside the H$_2$O ice line, whereas $\sim 5$ \% originated from beyond the CO$_2$ ice line. In contrast, ice at $r=15$ AU almost exclusively originates from the region between the H$_2$S and CO ice lines at $r=5.6$ AU and $r=23$ AU, respectively.

\begin{figure*}[ht!]
    \centering
    \includegraphics[width=\textwidth]{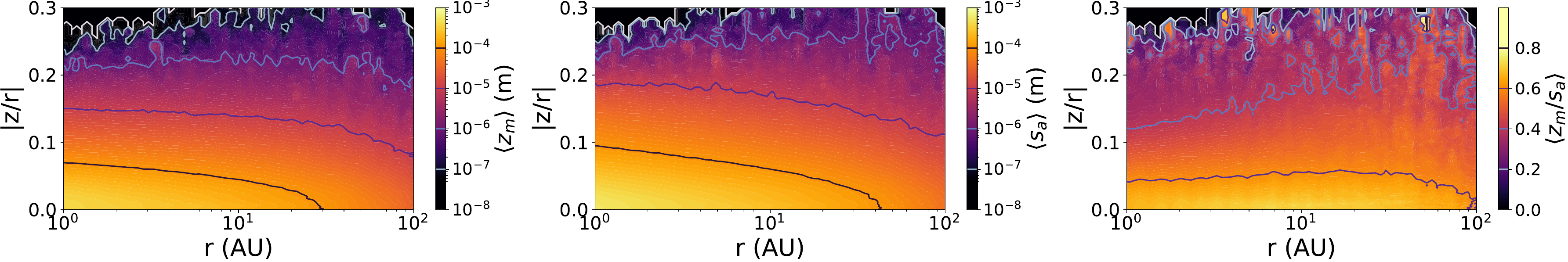}
    \caption{Expected values, calculated with Eq. \ref{eq:weighting}, for the monomer depth $z_\text{m}$ (left), home aggregate size $s_\text{a}$ (center), and $z_\text{m}/s_\text{a}$ (right) as a function of radial and vertical position throughout the sampling region. The contours in all three panels highlight specific values of the quantities depicted.}
    \label{fig:32NonLocalF3_cleaned3}
\end{figure*}
\begin{figure}[ht!]
    \centering
    \includegraphics[width=.45\textwidth]{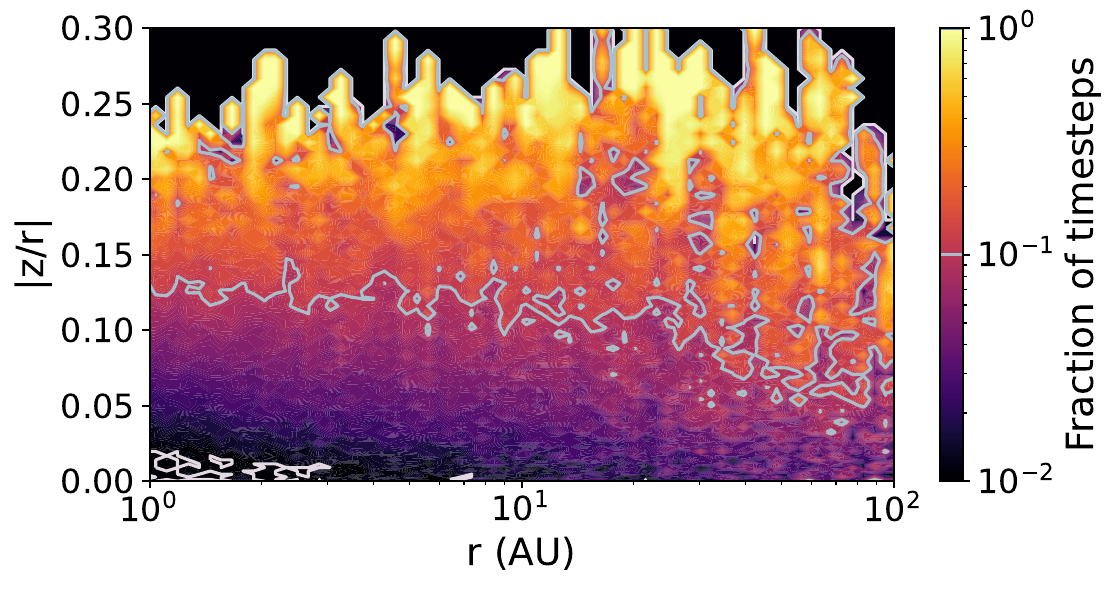}
    \caption{Weighted fraction of timesteps that monomers spend exposed to the gas phase as a function of radial and vertical position in the disk.}
    \label{fig:32NonLocalF3_cleaned3Exposedness}
\end{figure}

%% file: TextFiles/Section3/Section32.tex
\begin{figure*}[ht!]
    \centering
    \includegraphics[width=.95\textwidth]{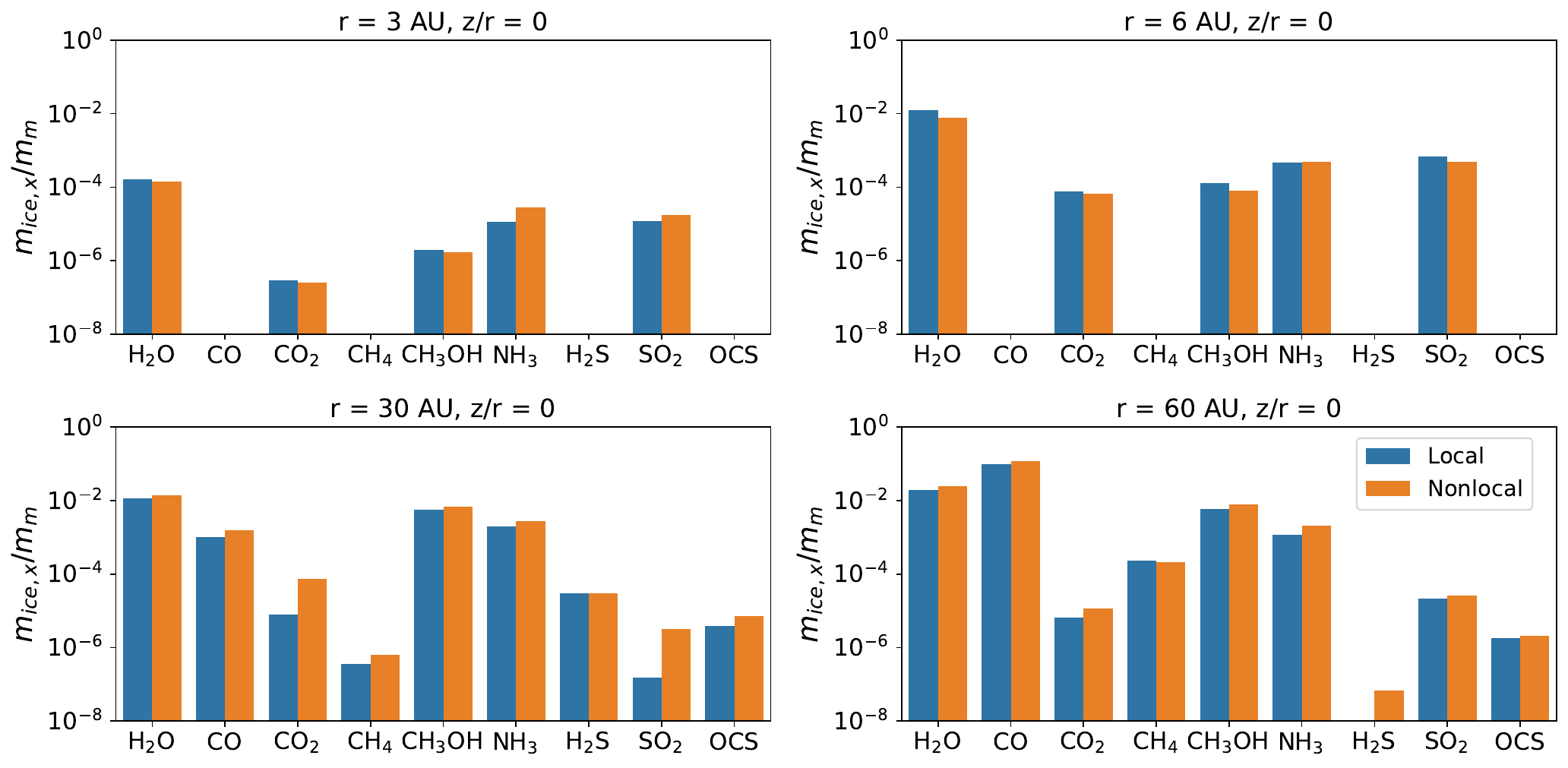}
    \caption{Comparison of the expected ice mass in units of monomer mass of the various species between the nonlocal population and a local monomer population evaluated fixed at their respective positions.}
    \label{fig:33NonLocalF3_cleaned3NonlocalHistograms}
\end{figure*}
\begin{figure}[ht!]
    \centering
    \includegraphics[width=.50\textwidth]{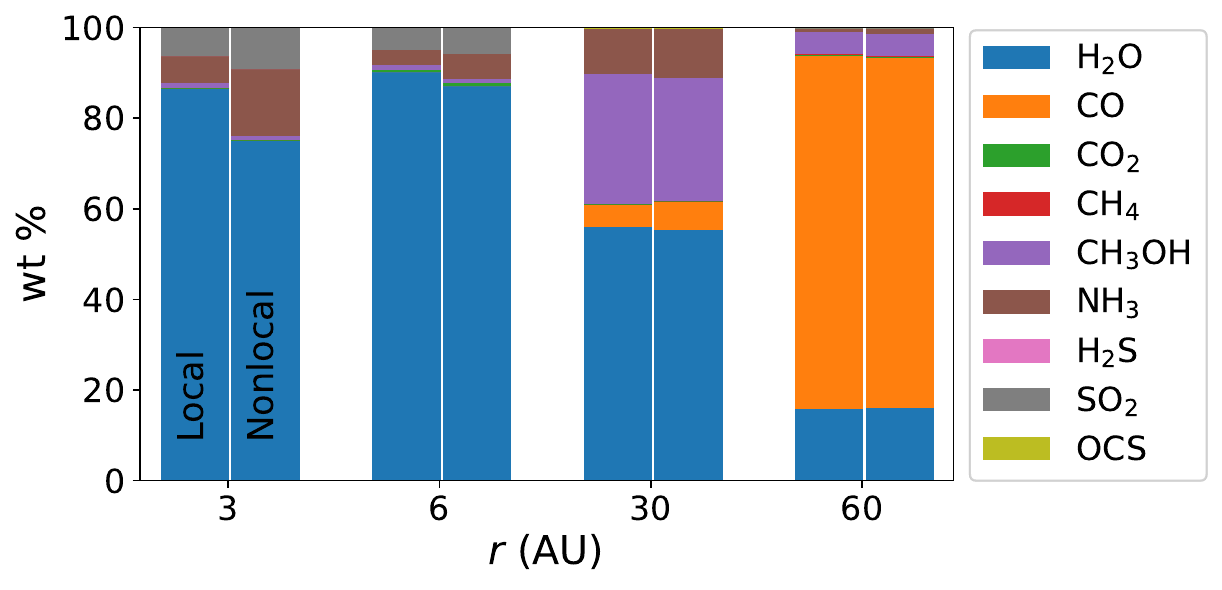}
    \caption{Expected monomer ice composition in the midplane at $r=3,6,30$, and $60$ AU for the amounts of ices shown in Fig. \ref{fig:33NonLocalF3_cleaned3NonlocalHistograms} for local and nonlocal monomers.}
    \label{fig:33NonLocalF3_cleaned3NonLocalComposition}
\end{figure}
The results from Sect. \ref{sec:3.1} show that significant fractions of a given local dust population can originate from disk regions that are chemically distinct from the local disk region of the dust population. The ice budget of monomers can also be significantly processed during the journey from their initial position to the location of the assessed dust population. The extent of processing primarily depends on whether the monomer is exposed to the gas phase, which allows monomers to adsorb molecules from disk regions which can be chemically distinct from their region of origin. This would make the final ice composition of the monomer nonlocal. If the monomer is shielded from the gas phase, the monomer can only undergo thermal desorption, which results in ice of different species to be lost at their respective ice lines. In that case the chemical composition of the monomer remains relatively unaffected by the varying chemical composition of the gas phase at different locations in the disk.\\
Whether a monomer is exposed to the gas phase is determined by the monomer depth $z_\text{m}$. However, the monomer depth depends on the collisional history of the monomer, in which the home aggregate size $s_\text{a}$ plays an important role. We therefore explore the expected values for the aggregate size and monomer depth throughout the disk in Fig. \ref{fig:32NonLocalF3_cleaned3}. Both quantities appear to follow a similar pattern, with both the average monomer depth and aggregate size decreasing as a function of $r$ and $z$. The latter is expected since monomers spend most time in aggregates close to the local maximum grain size $a_\text{max}$, as those aggregates contain most mass. Furthermore, the correlation between $s_\text{a}$ and $z_\text{m}$ is a consequence of the fact that coagulation is the dominating collision outcome in our collision model, which means that monomers are preferentially buried in larger aggregates. In contrast, erosion and fragmentation reset the monomer depth to a value drawn from a spherically uniform distribution, and thus tend to position monomers closer to the aggregate surface \citep{Oosterloo+2023}. The rightmost panel of Fig. \ref{fig:32NonLocalF3_cleaned3} shows the expected relative monomer depth, $\langle z_\text{m}/s_\text{a}\rangle$, which increases as a function of $r$ up to the disk region between 10 AU and 20 AU. This is a consequence of decreasing collision rates as a function of $r$, which means that in the integration time of 100 kyr, erosion and fragmentation events may be so rare that they may not happen at all. Therefore, monomers are preferentially buried by coagulation even more than at smaller $r$. For $r\gtrsim20$ AU, the relative monomer depth decreases, which is a consequence of the more rapidly decreasing maximum grain size $a_\text{max}$  at large $r$, which may in turn result in more frequent erosion and fragmentation \citep[c.f. Fig. 2 from][]{Oosterloo+2023}{}.\\
Another trend visible in Fig. \ref{fig:32NonLocalF3_cleaned3} is a decrease in $\langle s_\text{a} \rangle$ and $\langle z_\text{m}\rangle$ as a function of $z$. For the average aggregate size $\langle s_\text{a} \rangle$ this is expected as settling causes larger aggregates to be preferentially residing closer to the disk midplane, which automatically allows for larger values of $z_\text{m}$. Fig. \ref{fig:32NonLocalF3_cleaned3} also shows that $\langle z_\text{m}/s_\text{a}\rangle$ decreases as a function of $z$. Monomers with high $\langle z_\text{m}/s_\text{a}\rangle$ have been buried by a large number of collisions resulting in aggregate growth. However, aggregates that have undergone significant collisional growth usually have larger $s_\text{a}$ and thus tend to reside closer to the midplane, selectively removing monomers with higher $z_\text{m}/s_\text{a}$ from higher $z$.\\
Fig. \ref{fig:32NonLocalF3_cleaned3Exposedness} shows the weighted fraction of timesteps monomers spend exposed, and thus allowed to undergo adsorption and photodesorption. We here use Eq. \ref{eq:weighting}, and set $X_{nm}=1$ for exposed data points, and $X_{nm}=0$ for unexposed data points. Since the timestep size $\Delta t$ is one of the weights in Eq. \ref{eq:weighting}, $w_{\Delta t}$, the weighted number of timesteps is equivalent to the fraction of time monomers are exposed, weighted by $w_p$ and $w_g$ in this context (see Sect. \ref{sec:2.3}). Between 1 and 100 AU, most monomers below $z/r=0.05$ spend less than $10\%$ of the time exposed to the gas phase. In the grid cells closest to the midplane, we find that between $r=1$ AU and $r=5$ AU, monomers are exposed for less than $1\%$ of their time. This low fraction of exposed monomers also explains the significant amount of noise in Fig. \ref{fig:32NonLocalF3_cleaned3Exposedness} compared to Fig. \ref{fig:32NonLocalF3_cleaned3}. In the calculation of the fraction of timesteps monomers are exposed, only the exposed data points are used in Eq. \ref{eq:weighting}, significantly limiting the number of data points per grid cell in Fig. \ref{fig:32NonLocalF3_cleaned3Exposedness} compared to other figures calculated with Eq. \ref{eq:weighting}.\\
Altogether Fig. \ref{fig:32NonLocalF3_cleaned3Exposedness} demonstrates that throughout most of the sampling region, most monomers are buried well below the aggregate surface. These monomers are only able to lose ice via thermal desorption, while monomer ice processing occurs over a comparatively short time compared to the entire monomer history.

%% file: TextFiles/Section3/Section33.tex
As a next step, we explore the effects of the dynamical and collisional processing on the ice composition on local dust. In order to quantify the effects of nonlocal disk processes of the ice, we compare the expected amount of ice predicted at various positions in the nonlocal simulation to a set of local monomer simulations without transport processes that are otherwise identical to the nonlocal monomers (i.e. $r$ and $z$ remain fixed for all monomers). At each position we evolve 8000 unique, local monomers, a number comparable with the number of unique nonlocal monomers visiting each grid cell in the disk midplane (c.f. Fig. \ref{fig:ADNonLocalF3_cleaned3DataPoints}).\\
Fig. \ref{fig:33NonLocalF3_cleaned3NonlocalHistograms} and Fig. \ref{fig:33NonLocalF3_cleaned3NonLocalComposition} present this comparison at $r=3,6,30,60$ AU in the midplane. Fig \ref{fig:33NonLocalF3_cleaned3NonlocalHistograms} depicts the resulting absolute amounts of individual ice species, while Fig. \ref{fig:33NonLocalF3_cleaned3NonLocalComposition} shows the overall ice compositions. It becomes clear that at 3 AU and 6 AU, H$_2$O is the dominant ice phase molecule in both the local and nonlocal case, making up $\gtrsim 75\%$ of the ice mantle in all cases. At 30 AU, we find in both cases an average ice mantle that is dominated by H$_2$O and CH$_3$OH, with smaller amounts of CO and NH$_3$. At 60 AU, CO has become the dominant ice molecule, making up $\sim 78\%$ of the average monomer ice mantle in both the local and nonlocal population. However, H$_2$O also still makes up a significant fraction ($\sim16\%$) of the average monomer ice mantle.\\
Despite the comparable ice composition in terms of major species in Fig. \ref{fig:33NonLocalF3_cleaned3NonLocalComposition}, it is evident from both Fig. \ref{fig:33NonLocalF3_cleaned3NonlocalHistograms} and Fig. \ref{fig:33NonLocalF3_cleaned3NonLocalComposition} that nonlocal disk processing can have a significant effect on the amount of volatile ice of individual chemical species. The precise effect of nonlocal disk processing on individual ice species is position-dependent, with its effect on bulk ice composition being more clear at $r=3$ and $r=6$ AU. For most ice species, the amounts of ice on local and nonlocal monomers usually differ by up to a factor of two. Notable exceptions are CO$_2$ and the sulfur-bearing molecules, which are explored in more detail in Sect. \ref{sec:3.4}. Here, we note the presence of CO$_2$ as ice at $r=3$ AU in Fig. \ref{fig:33NonLocalF3_cleaned3NonlocalHistograms}, in particular since the CO$_2$ ice line is located at 5 to 10 AU, depending on ice line definition (c.f. Fig. \ref{fig:41DiffusionStudy}). This is due to the CO$_2$ molecules residing in an ice mantle dominated by H$_2$O and NH$_3$. In addition, closer inspection revealed that the number of molecules on the monomer is one order of magnitude larger than the number of molecule sites available for thermal desorption \citep{Aikawa+1996, Woitke+2009, Oosterloo+2023}. This significantly reduces the desorption rate for CO$_2$ with respect to a pure CO$_2$ ice mantle. Since adsorption is not affected by the ice mantle structure in our model, and is much larger than at the CO$_2$ ice line due to the high abundance of CO$_2$ in the gas phase at 3 AU, a small amount of CO$_2$ is able to exist as a trace ice at $r=3$ AU.  

%% file: TextFiles/Section3/Section34.tex
\begin{figure*}[ht!]
    \centering
    \includegraphics[width=.99\textwidth]{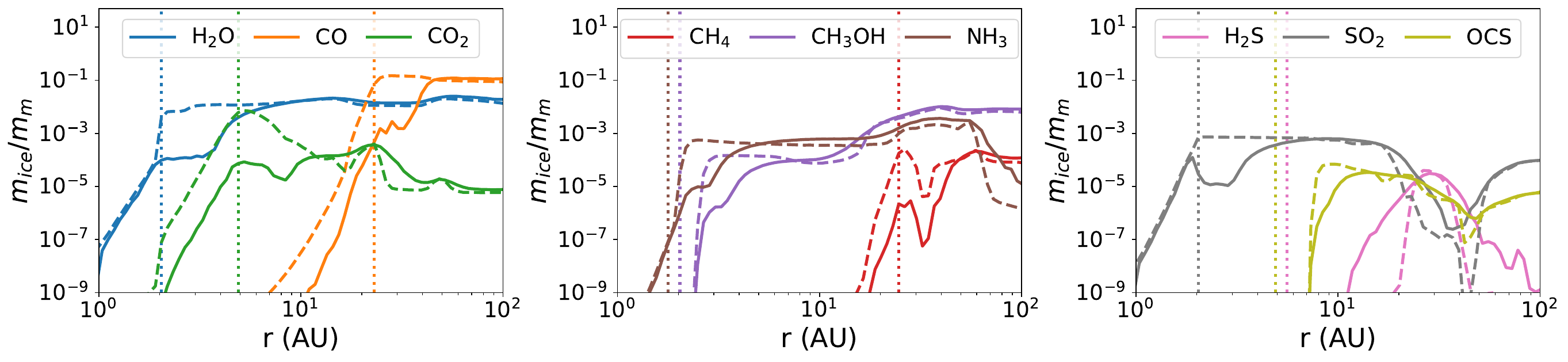}
    \caption{Comparison of the midplane abundance of key ice species predicted by SHAMPOO (solid line) and estimated from the local ice phase molecule number density $n_{x,\text{ice}}$ in ProDiMo (dashed line) for H$_2$O, CO, and CO$_2$ (left); CH$_4$, CH$_3$OH, and NH$_3$ (center); and H$_2$S, SO$_2$, and OCS (right). The vertical dotted lines indicate ice lines of the respective species.}
    \label{fig:34MidplaneIcesComparison}
\end{figure*}
\begin{figure*}[ht!]
    \centering
    \includegraphics[width=.84\textwidth]{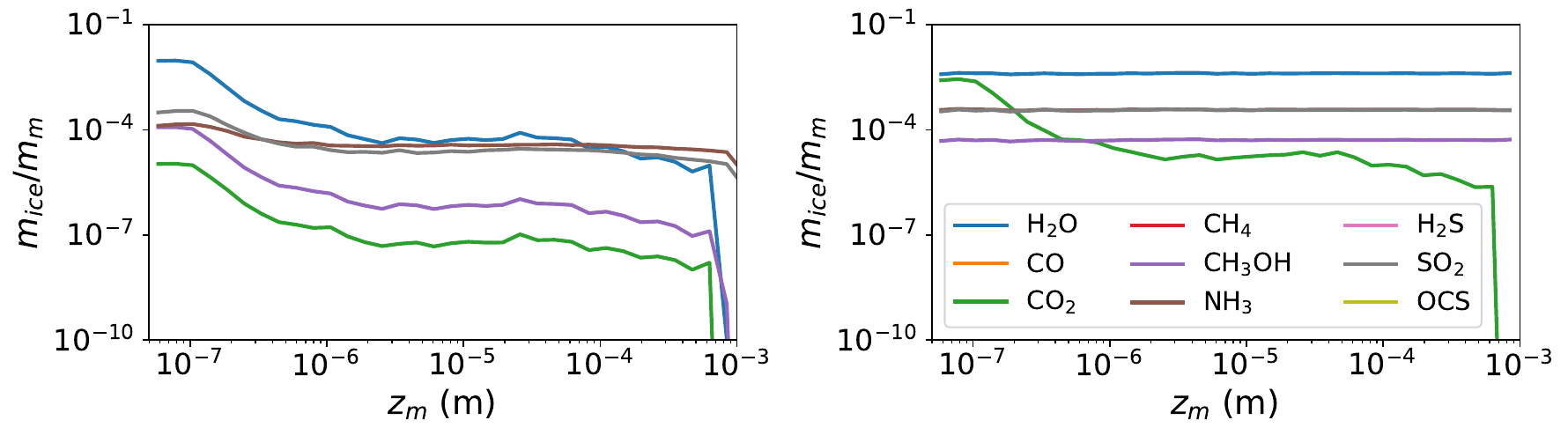}
    \caption{Amount of ice on monomers as a function of monomer depth $z_m$ at $r=3$ AU (left), and $r=5$ AU (right) in the region below $|z/r|=0.1$.}
    \label{fig:34NonLocalF3_cleaned3AggregateIceDistributions}
\end{figure*}
\begin{figure*}[ht!]
    \centering
    \includegraphics[width=.99\textwidth]{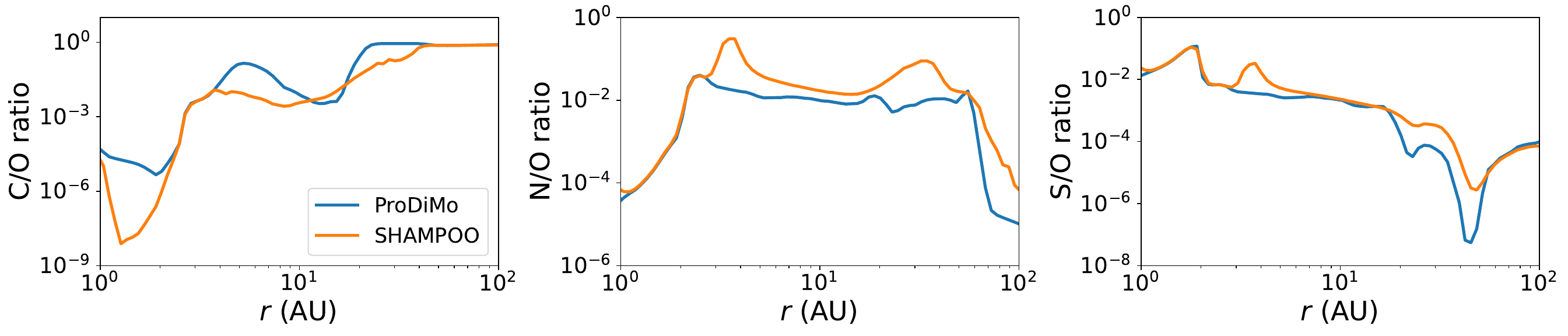}
    \caption{Comparison of the C/O (left), N/O (center), and S/O (right) elemental ratios in the ice mantles of nonlocal monomers in SHAMPOO and locally processed dust in ProDiMo.}
    \label{fig:35MidplaneRatioComparison}
\end{figure*}

For a more comprehensive understanding of the behavior of ice on nonlocal monomers we consider the expected amount of ice for the nonlocal monomers as a function of $r$ in the disk midplane in Fig. \ref{fig:34MidplaneIcesComparison}. In order to contextualize the behavior of nonlocal dust, we compare our results to the expected amount of ice in the ProDiMo background disk model, which treats ice evolution as a fully local process. We here estimate the average amount of ice per monomer mass from ProDiMo via the local ice number density $n_{x,\text{ice}}$ \citep[][]{Oosterloo+2023}{}{}:
\begin{align}
\label{eq:initIce}
    \frac{m_{x,\text{ice}}}{m_\text{m}}\Biggr\rvert_\text{ProDiMo}=\frac{n_{x,\text{ice}}m_x}{\rho_\text{d}}.
\end{align}
Here, $m_x$ denotes the molecular mass of species $x$, and $\rho_\text{d}$ the local dust mass density.\\
From Fig. \ref{fig:34MidplaneIcesComparison}, gradients in the amount of ice appear smoother in SHAMPOO compared to ProDiMo. Examples of this smoother distribution of ice can be seen for CO$_2$ between 10 and 100 AU and beyond 6 AU for CH$_3$OH and NH$_3$. For the sulfur-bearing species, this behavior is also visible between 10 and 100 AU. We note that for all the above species, the highlighted regions are significantly behind the ice lines of the respective species, which means that the timescales associated with adsorption and desorption are usually much longer than the timescales associated with dynamical processes \citep{Oosterloo+2023}. Therefore, this smoothing effect is a result of diffusion, which is the dominant dynamical process throughout most of the disk (c.f. Fig. \ref{fig:AENonLocalF3_cleaned3StokesNumbers}). This also means that species with originally large spatial variations in ice abundance are most affected. Indeed, Fig. \ref{fig:34MidplaneIcesComparison} and Fig. \ref{fig:AABackgroundModelAbundances} show that this is mostly the case for CO$_2$ and the sulfur-bearing volatiles. This also explains the larger differences seen for these specific molecules in Fig. \ref{fig:33NonLocalF3_cleaned3NonlocalHistograms} between local and nonlocal monomer populations. Another indication of diffusion is the significantly larger amount of H$_2$S ice being located between 10 and 20 AU in our model compared to ProDiMo. Although the ice line of H$_2$S is located around 6 AU, there is almost no H$_2$S present in either the gas or the ice phase between 6 and 20 AU in the background model (c.f. Fig. \ref{fig:AABackgroundModelAbundances}). Although monomers originating from between 6 and 20 AU initially contain no H$_2$S, this region is allowed to become partially enriched in H$_2$S due to the influx of monomers from beyond 20 AU, where significant amounts of H$_2$S can be accumulated as ice. \\
The largest differences between the amount of ices predicted by our model and by ProDiMo occur for the amounts of ice immediately behind ice lines. At these locations, our model predicts up to two orders of magnitude less ice for specific molecules than ProDiMo. Although we note this behavior for all molecules considered except H$_2$S, it is most clearly visible for H$_2$O, CO$_2$, CO and SO$_2$. This can be related to the systematic distribution of ice as a function of monomer depth $z_\text{m}$.\\
In order to explore this, we consider all monomers located below $|z/r|=0.1$ to ensure a sufficient number of data points at every bin of $z_\text{m}$. Although the size distribution of aggregates varies as a function of vertical $z$ due to settling, the thermal and chemical structure remains approximately constant below this height. The resulting distributions found at $r=3$ AU and $r=5$ AU are shown in Fig. \ref{fig:34NonLocalF3_cleaned3AggregateIceDistributions}.\\
At $r=3$ AU (left panel of Fig. \ref{fig:34NonLocalF3_cleaned3AggregateIceDistributions}), monomers located at shallower depth than the critical monomer depth $z_\text{crit}$ contain more than two orders of magnitude more H$_2$O, CH$_3$OH and CO$_2$ ice than those located at larger $z_\text{m}$, while the difference appears to be smaller for NH$_3$ and SO$_2$. Monomers located at shallower depth than $z_\text{crit}$ are always exposed to the gas phase in our model, and are thus allowed to undergo adsorption. At greater monomer depth, the probability for a monomer to be exposed drops rapidly, becoming effectively zero for monomers located deeper than $z_\text{m}\sim 3\e{-7}$ m \citep[see][]{Oosterloo+2023}. Unexposed monomers can lose ice through thermal desorption but cannot accumulate new ice. Although the aggregate is strictly speaking in the ice-forming region, the thermal desorption rate is still significant in the region immediately behind the ice line. This means that unexposed monomers will gradually lose ice over time in this disk region. Therefore, monomers buried deeper in aggregates can only have significant amounts of ice if they are regularly exposed to the gas phase on a timescale comparable to or shorter than the thermal desorption timescale. Since the dust mass in the midplane is dominated by unexposed monomers, this \textit{microscopic collisional mixing} results in a significantly lower amount of ice per unit dust mass with respect to ProDiMo, which does not couple collisional processing and ice evolution.\\
It should be noted that this result was obtained in a context where thermally desorbed molecules are assumed to escape the home aggregate instantaneously. In practice, finite timescales for readsorption of molecules on other monomers in the aggregate and diffusion likely slow down the escape of molecules from the aggregate interior, which means that the average amounts of ice shown in Fig. \ref{fig:34MidplaneIcesComparison} represent lower limits for the various amounts of ice (see also Sect. \ref{sec:4.1}).\\
At $r=5$ AU it appears that the above effect has disappeared for all ice species except CO$_2$. This can be traced back to the fact that CO$_2$ is the most volatile species present in appreciable amounts at $r=5$ AU, just behind the ice line for CO$_2$ (c.f. Fig. \ref{fig:31NonLocalF3_cleaned3ChemicalOrigin}). This means that CO$_2$ is still undergoing considerable thermal desorption at $r=5$ AU, whereas both the adsorption rate and desorption rate have become negligible over 100 kyr timescales for the other species, whose ice lines are located around $r=2$ AU.

%% file: TextFiles/Section3/Section35.tex
The ice masses shown in Fig. \ref{fig:34MidplaneIcesComparison} give rise to the radial profiles of C/O, N/O and S/O elemental abundance ratios shown in Fig. \ref{fig:35MidplaneRatioComparison}. These are compared with the ratios predicted by the background ProDiMo model at the same positions.\\
The lower amounts of major volatile carbon-bearing species have a significant effect on the C/O ratio, which is governed by the relative amounts of H$_2$O, CO, CH$_3$OH and CO$_2$ ices. In particular between $r=1$ and $r=2$ AU, our model results in an even lower C/O ratio than the ratio resulting from ProDiMo ($\sim 10^{-8}$ compared to $\sim 10^{-5}$). This is due to the absence of any significant amounts of volatile carbon-bearing ices at these radial distances. Although our model predicts similar amounts of H$_2$O-ice (the dominant oxygen-carrier in this region), as ProDiMo, our model predicts less volatile carbon-bearing ices at these radial positions due to collisional mixing, resulting in a lower C/O ratio. As a function of radial position, our model and ProDiMo are in better agreement between $r=2$ and $r=4$ AU, the region where all major carbon- and oxygen-bearing ices present in significant amounts (H$_2$O, CH$_3$OH and CO$_2$) are affected by collisional mixing in our model. It becomes clear that in this disk region, collisional mixing does only affect the total amount of carbon- and oxygen-bearing ices, but does not significantly affect the C/O ratio. In the region between $r=4$ AU and $r=10$ AU, where our model predicts significantly less CO$_2$-ice, resulting in a lower C/O ratio compared to ProDiMo. A similar effect is also visible for CO between $r\eqsim 20$ AU and $r\eqsim 40$ AU. Overall, it appears that the predicted C/O ratio in our model is usually significantly lower than in ProDiMo.\\
The lower amount of major volatile carbon-bearing species also affects the N/O and S/O ratio as CO and CO$_2$ also make significant contributions to the volatile oxygen reservoir, which results in higher N/O and S/O ratios with respect to ProDiMo near the CO$_2$ and CO ice lines. Exterior to $r=40$ AU, our model predicts amounts of ice similar to ProDiMo for all major carbon- and oxygen-bearing species, which results in the two models being in agreement on the C/O ratio at these distances. For the N/O ratio, we also note that the decrease for $r>50$ AU in the N/O ratio inferred from ProDiMo, which is associated with a strong decrease in the amount of NH$_3$ ice. It is clear that the shallower gradient in NH$_3$ predicted by our model originating from diffusion in Fig. \ref{fig:34MidplaneIcesComparison} also translates into a shallower gradient in the predicted N/O ratio in our model. The S/O ratio also exhibits significant dips at $r\eqsim 25$ AU and $r=40$ AU, which can be fully traced back to the condensation behavior of SO$_2$ and H$_2$S in Fig. \ref{fig:34MidplaneIcesComparison}, which significantly shapes the total budget of volatile sulfur. Again, diffusion results in a smoothing of these features in our model with respect to ProDiMo.

%% file: TextFiles/4Discussion.tex
\label{sec:4}

\subsection{Molecule diffusion inside aggregates}
\label{sec:4.1}
\input{TextFiles/Section4/Section41.tex}
\subsection{Comparison with other dust evolution models}
\label{sec:4.2}
\input{TextFiles/Section4/Section42.tex}

\subsection{Implications for CHNOS budgets of planetesimals}
\label{sec:4.3}
\input{TextFiles/Section4/Section43.tex}

%% file: TextFiles/Section4/Section41.tex
\begin{figure*}[ht!]
    \centering
    \includegraphics[width=.99\textwidth]{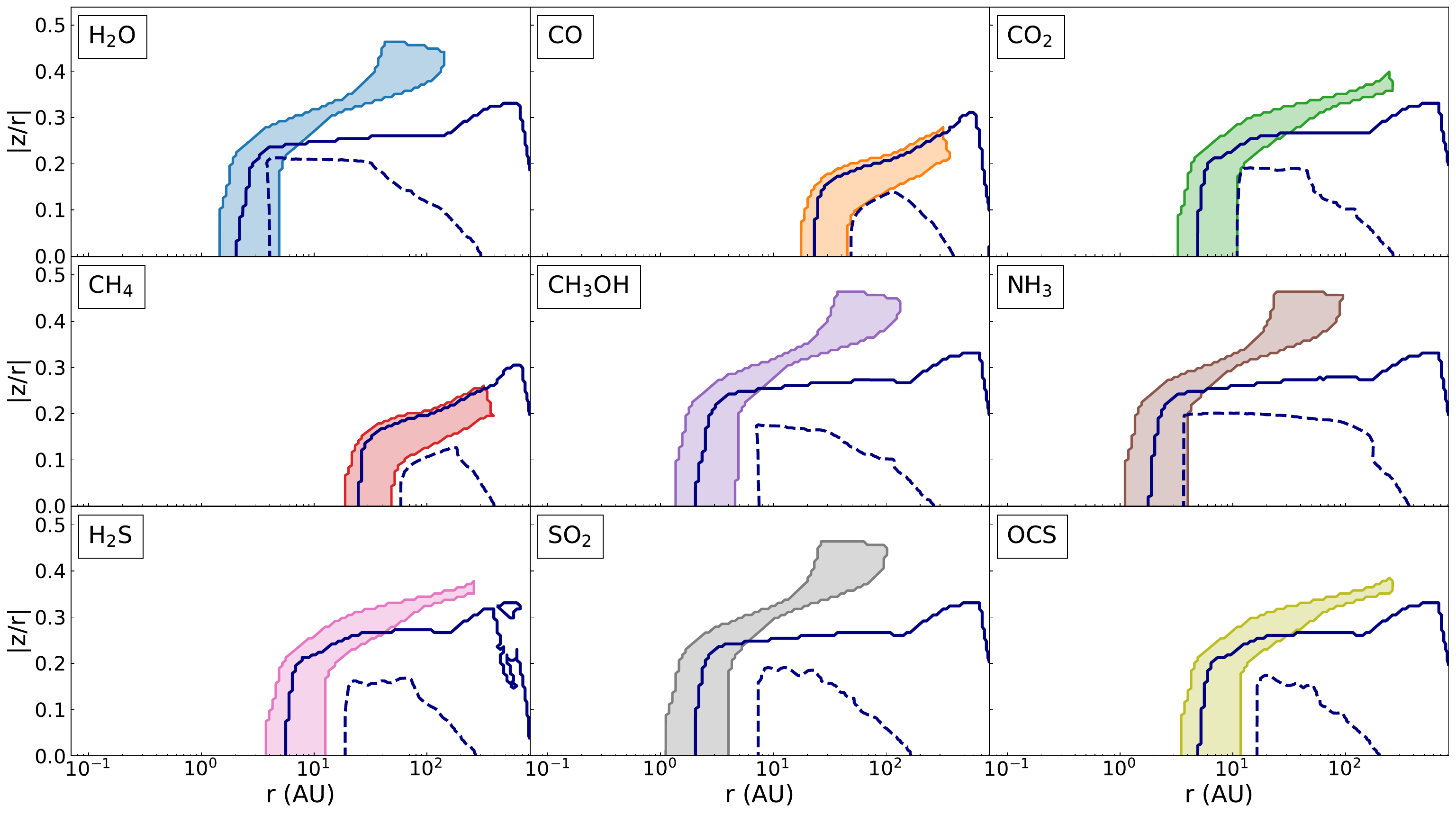}
    \caption{Map of the spatial zone where diffusion is expected to significantly affect the ice evolution of monomers for the eight chemical species considered throughout this work. The dark blue lines indicate the ice line of the respective species informed from either the 50\% condensation threshold in the background model (solid) or the location where the adsorption and desorption rates are equal (dashed).}
    \label{fig:41DiffusionStudy}
\end{figure*}

The results presented in Sect. \ref{sec:3.3}-\ref{sec:3.5} all pertain to monomers which are only allowed to undergo adsorption and photodesorption when located at a monomer depth near or below the critical monomer depth $z_\text{crit}$, while thermal adsorption is assumed to be efficient in removing ice from monomers throughout the entire aggregate. In reality, however, some of the molecules may be recaptured by other monomers in the same aggregate through readsorption, which would lead to the current model underestimating the amount of ice on monomers \citep{Oosterloo+2023}. In this section we aim to spatially constrain the regions in the background disk model where the effects of re-adsorption could be significant via a timescale analysis.\\
For this purpose, we aim to derive a timescale for which significant amounts of ices of species $x$ can escape an aggregate of given filling factor $\phi$ and size $s_\text{a}$. As a starting point, we will here use that the three-dimensional root mean square displacement $\langle r^2 \rangle$ after time interval $t$ associated with particles undergoing diffusion from a point source at $r=0$ can be expressed as
\begin{align}
\label{eq:diffusion}
    \langle r^2 \rangle=6Dt.
\end{align}
Here, the diffusivity $D=\lambda v_0$ depends on the molecule mean free path $\lambda$ and diffusion velocity $v_0$. If molecules only collide with monomers inside the aggregate and monomers are distributed homogeneously throughout the aggregate, we can express the mean free path inside an aggregate as in \cite{Oosterloo+2023}
\begin{align}
    \lambda = \frac{4s_\text{m}}{3\phi}.
\end{align}
Here $s_\text{m}$ is the monomer radius and $\phi$ the home aggregate filling factor. The diffusion velocity $v_0$ can be expressed in terms of average time required for a molecule to traverse the mean free path $\lambda$ inside an aggregate \citep[see][]{Oosterloo+2023}:
\begin{align}
    v_0=\frac{\lambda}{\tau_\text{mc}+S\tau_\text{d}}.
\end{align}
Here, $\tau_\text{mc}$ denotes the molecule collision timescale, $S$ the sticking factor and $\tau_\text{d}$ the molecule desorption timescale, the average time a molecule resides on the surface of a monomer before desorbing. Expressions for $\tau_\text{mc}$ and $\tau_\text{d}$ have been derived by \cite{Oosterloo+2023}, 
\begin{align*}
    \tau_\text{mc}=\frac{\lambda}{d_\text{s}\nu_x}\exp\left(\frac{E_{\text{ads},x}}{2k_\text{B}T_\text{d}}\right),\quad \tau_\text{d}=\frac{1}{\nu_x}\exp\left(\frac{E_{\text{ads},x}}{k_\text{B}T_\text{d}}\right).
\end{align*}
Here $\nu_x$ denotes the monomer ice lattice vibration frequency, $d_\text{s}$ the average distance between two molecule binding sites in the monomer ice lattice. We can write the molecule diffusivity as
\begin{align}
    D=\lambda^2\nu_x\left[\frac{\lambda}{d_\text{s}}\exp\left(\frac{E_{\text{ads},x}}{2k_\text{B}T_\text{d}}\right)+S\exp\left(\frac{E_{\text{ads},x}}{k_\text{B}T_\text{d}}\right)\right]^{-1}.
\end{align}
In the context of readsorption, we can thus define the \textit{diffusion timescale} $\tau_\text{diff}$ using Eq. \ref{eq:diffusion}:
\begin{align}
\label{eq:taudiff}
\tau_\text{diff}\left(\langle r^2 \rangle\right)&=\frac{\langle r^2 \rangle}{6D}\notag\\
&=\frac{\langle r^2 \rangle}{6\lambda^2\nu_x}\left[\frac{\lambda}{d_\text{s}}\exp\left(\frac{E_{\text{ads},x}}{2k_\text{B}T_\text{d}}\right)+S\exp\left(\frac{E_{\text{ads},x}}{k_\text{B}T_\text{d}}\right)\right].
\end{align}
The importance of diffusion depends on how the diffusion timescale compares to the timescale of our simulations ($\sim 100$ kyr) and the timescales of other disk processes. In this context we can define an upper and lower bound on $\tau_\text{diff}$ where diffusion is relevant. We define the upper bound of the diffusion timescale as the threshold where the time required for the typical molecule in an aggregate to diffuse one mean free path is ten times longer than the simulation timescale ($\tau_\text{diff}(\lambda^2)> 10^6\,\unit{yr}{}$). Similarly, we define the lower bound of the diffusion timescale as the condition that the typical molecule in an aggregate is able to exit an aggregate of the most probable dust aggregate size $\Bar{s}_\text{a}$\footnote{The most probable aggregate size $\Bar{s}_\text{a}$ is obtained from the local dust size distribution specified in the background disk model, and should not be confused with $\langle s_\text{a}\rangle $, the expected aggregate size.} within one year ($\tau_\text{diff}(\Bar{s}_\text{a}^2)<1\,\unit{yr}{}$). This is comparable or shorter than any of the dynamical timescale and collisional timescales at any location where $r>1$ AU \citep[see][]{Oosterloo+2023}. At some locations throughout the disk, the most probable aggregate size is smaller than the mean free path within an aggregate. This can be particularly true high in the disk atmosphere, where dust grains are very small. In this case collisions between molecules and monomers in the aggregate become rare, as most molecules would escape the aggregate right away upon desorption.\\
Altogether the disk region where diffusion within an aggregate may play a role is limited by the above three boundary conditions defined in terms of $\tau_\text{diff}$, $\lambda$ and $\Bar{s}_\text{a}$. Calculation of $\tau_\text{diff}$ and $\Bar{s}_\text{a}$ throughout the entire background disk model allows for mapping out the disk regions for different species. Figure \ref{fig:41DiffusionStudy} highlights these different regions for the nine chemical species considered in this work. In these regions, none of the above three conditions is true for the respective species, and thus diffusion can play a role in setting the ice budgets of individual monomers over 100 kyr. We also compare the locations of these regions with the ice lines of respective species. For each species, Fig. \ref{fig:41DiffusionStudy} shows two different ice line positions, derived from different definitions. In the first definition, the ice line of a given species can be calculated from the 50 \% condenstion threshold in the background model (the location where 50 \% of all molecules of a given species exist as ice). The second ice line definition is derived from the equivalence of the specific adsorption and total desorption rates ($\mathcal{R}_{\text{ads},x}=\mathcal{R}_{\text{tds},x}+\mathcal{R}_{\text{pds},x}$). It becomes clear that up to $z/r\sim 0.2-0.3$, the region where diffusion may affect monomer ice budgets on the timescales considered in this work is tied to the position of the ice line. Higher in the disk atmosphere, photodesorption starts to significantly impact the desorption of ice molecules due to higher background UV flux, limiting the vertical height of the ice-dominated region. This does not directly affect diffusion within the aggregate, since $\tau_\text{diff}$ only depends on the local dust temperature $T_\text{d}$ in the background model, which is only affected indirectly by the higher UV flux \citep{Woitke+2009}. Therefore, the diffusion region is vertically more extended than the ice-dominated regions for most species, with the exception of species with very low $E_\text{\text{ads},x}$ such as CO and CH$_4$. However, for species with higher $E_\text{\text{ads},x}$, such as H$_2$O, CH$_3$OH, NH$_3$ and SO$_2$, the diffusion region is also vertically limited, albeit to higher $z/r\sim 0.4-0.5$. This is in part an effect from the increasing dust temperature, although $\Bar{s}_\text{a}$ also decreases as a function of $z$, causing the mean free path to exceed $\Bar{s}_\text{a}$ at these large heights.\\
Altogether it becomes clear that in the disk regions most relevant for planet formation, diffusion within the aggregate could have the largest effect on the amount of ice of individual species around their respective ice lines, allowing for more ices to be retained in aggregate interiors than reported in this study. Radially, this region can extend several AU's up to tens of AU's for more volatile species. Vertically, the extension of the region where diffusion may affect our results is also shaped by the volatility of the species, with the region being contained closer to the midplane (as low as $z/r\eqsim 0.2$) for volatile species, while comparatively less volatile species do have a vertically more extended region. 

%% file: TextFiles/Section4/Section42.tex
The model that is conceptually most similar to SHAMPOO is the dust evolution model presented in \cite{Krijt+2016b}. The collision model used in SHAMPOO is identical to the collision model presented in \cite{Krijt&Ciesla2016}, and is also used in the context of ice processing in \cite{Krijt+2016b} to investigate the interplay of settling, turbulent mixing and collisional growth on the evolution of the amount of H$_2$O ice on dust grains. It was found in \cite{Krijt+2016b} that H$_2$O can be efficiently locked away in ice-rich aggregates near the disk midplane, while we do not discern such an effect in our model. This is likely the consequence of three effects. Firstly, \cite{Krijt+2016b} assume that all H$_2$O ice is situated at the aggregate surface, whereas this does not have to be the case in our model. Since escape of molecules from the aggregate interior is assumed to be efficient in our model, significant amounts of H$_2$O can be released back into the gas phase in our model. Secondly, our model does not account for the effects of H$_2$O on the sticking properties of dust grains. Dust grains rich in H$_2$O ice can have fragmentation velocities up to an order of magnitude larger than pure silica grains \citep{Wada+2013, Gundlach&Blum2015}. This allows dust grains to grow to sizes of $s_\text{a}\eqsim 10^{-1}$ m within a $10$ kyr timescale in \cite{Krijt+2016b}. In the model of \cite{Krijt+2016b}, these aggregates have significant Stokes numbers, and thus a large fraction of the dust mass remains trapped in large aggregates in the disk midplane. Thirdly, the background disk model used in \cite{Krijt+2016b} is based on the minimum mass solar nebula \citep{Weidenschilling1977, Hayashi1981}, while our background disk model simulates a younger, more massive disk. Therefore, a dust grain of given size would have a lower Stokes number in our model due to the higher gas density, resulting in weaker settling of dust towards the disk midplane.\\
Another model conceptually similar to SHAMPOO is presented in \cite{Bergner+2021}, where an identical dynamical model is used. \cite{Bergner+2021} consider a more complex ice structure while not accounting for ice adsorption and collisional processing. They explore the preservation of interstellar ices and find that the destruction of these ices is generally inefficient, except for small (<10 $\mu$m) dust grains at $r \eqsim$ 20 or 150 AU, where ice is destroyed due to dust grains being mixed towards the disk surface regions through diffusion. Since small and large grains are coupled in our model through collisional processing, this ice depletion could potentially propagate towards larger aggregates. We do not find any evidence for systematic ice destruction in Fig. \ref{fig:34MidplaneIcesComparison} that is associated with the systematic destruction of ices in surface disk layers. However, our model includes adsorption, which allows ices lost to be replenished by new ice condensing from the gas phase. Furthermore, we do not discern between inherited ices and ices accumulated in situ, which means that the absence of large changes in the amount of ice does not exclude the systematic destruction of inherited ices. In order to constrain the degree of pristine ice loss, isotopic ratios such as D/H, $^{13}$C/$^{12}$C, $^{15}$N/$^{14}$N, and $^{18}$O/$^{16}$O could provide a useful proxy \citep[see e.g.][]{Cleeves+2014, Visser+2018, Oberg&Bergin2020}.\\ 
The collisional mixing of monomers throughout different aggregates was found to be an important cause for ice loss from aggregates in this work. This is a consequence of the efficient thermal desorption from the aggregate interior. We do however note that this process does not necessarily have to result in a net depletion of the affected species. A finite diffusion timescale can result in larger amounts of ice being retained in aggregate interiors through readsorption (see Sect. \ref{sec:4.1}), limiting the ice loss from the aggregate interior, in particular for compact aggregates. This could even allow for collisional mixing to result into a net gain of ice in the aggregate, where ice-rich monomers are efficiently stored away in the interiors of large aggregates. Altogether it is clear that near most ice lines, there exists a sizable disk region immediately behind the ice line where the collision timescale is comparable to the adsorption and desorption timescales. The results in this work show that in these regions, ice processing comprises entire aggregates due to ongoing collisional processing rather than being confined to dust aggregate surfaces.\\
We note that collisional mixing is not the only effect that may play a role in setting the volatile CHNOS budgets near ice lines. For example, \citep{Marseille&Cazaux2011} demonstrated that behind the H$_2$O ice line, there could exist a region where bare and icy grains coexist as a consequence of the lower adsorption energy for H$_2$O on bare grain surfaces compared to a H$_2$O ice surface. In this work, we approximate the adsorption energy of H$_2$O with a single value, whereas \cite{Marseille&Cazaux2011} consider a range of adsorption energies depending on the amount of clustering of H$_2$O molecules on the surfaces of dust grains. This effect would operate in the same disk region where collisional mixing affects ice evolution. Therefore, the combination of these two effects could lower abundances in H$_2$O ice even further since the continuous collisional mixing of monomers would preferentially expose monomers depleted or even devoid of H$_2$O ice to the gas phase, whose surfaces would initially have a lower binding energy for H$_2$O than their H$_2$O-ice covered counterparts.\\
The effect of radial drift on the position of the H$_2$O ice line is also explored together with the CO and CO$_2$ ice lines in \cite{Piso+2015}. However, in order for radial drift to play a significant role in the dynamical evolution of monomers in their home aggregates, aggregates are required to have a significant Stokes number, which was found to not be the case inside $r=100$ AU in the young, massive disk considered in this work (see Appendix \ref{sec:AE}). Altogether the effects of radial drift described in \cite{Piso+2015} may play a role in dust evolution for pebble-sized (i.e. St$\sim 1$) dust grains in older disks.

%% file: TextFiles/Section4/Section43.tex
It is not straightforward to assess the implications of this work for the CHNOS budgets of planetesimals. The implications discussed here pertain to the initial volatile budgets of the first generation of planetesimals formed from a smooth disk. Another consideration with regards to the implications of our results for the first generation of planetesimals is whether significant amounts of processing of the ice on dust from protoplanetary disks takes place during the formation of planetesimals and in planetesimals themselves.\\
The disk structure assumed in this work represents a young Class I disk which is relatively massive compared to Class II disks \citep{Andrews2020}. We also note that Class I disks are considerably more dynamic than their class II counterparts due to episodic accretion outbursts and the presence of a surrounding envelope \citep[e.g.,][]{Fischer+2023}{}{}. These factors can significantly impact the physical structure of the disk in this evolutionary stage. Specifically, the episodic outbursts could result in peaks in the photodesorption rate for exposed monomers, in particular at higher $z/r$. This would give rise to more frequent resets of monomer ice mantles, and thus in lower overall amounts of volatile CHNOS in planetesimal-forming dust. Altogether the effects of time-dependent luminosity and disk evolution merit further exploration in future studies. However, we do note a number of factors that may limit the effects of episodic peaks in photodesorption. Fig. \ref{fig:32NonLocalF3_cleaned3Exposedness} and Fig. \ref{fig:ADNonLocalF3_cleaned3DataPoints} demonstrate that the fraction of monomers exposed and located at higher $z/r$ is a small subset of the total monomer population. Also, in multiple Class I disks, gaps have been identified which are commonly associated with planet formation \citep[e.g.,][]{Sheehan+2018, Segura-Cox+2020}, which implies that planetesimals might form rather quickly in class I disks.\\
There are various mechanisms through which planetesimals could form \citep[see][for a more elaborate review]{Johansen+2014}. Planetesimals may form directly through pairwise collisions via high-velocity mass transfer collisions \citep{Teiser&Wurm2009, Windmark+2012, Garaud+2013} or porous growth \citep{Okuzumi+2012, Kataoka+2013}. However, the effectiveness of the former is disputed, and may take significantly longer than the timescales for planetesimal formation considered in this work \citep{Windmark+2012, Estrada+2016, BoothR+2018}{}. The latter can only occur for filling factors at least an order of magnitude lower than considered in this work. Therefore, the most likely formation channel for planetesimals in our modeling context is the concentration of dust by e.g. turbulent concentration or the streaming instability \citep{Cuzzi+2001, Hartlep&Cuzzi2020, Youdin&Goodman2005}. Previous modelling work has shown that no significant collisional processing takes place during the short final stage of the collapse of a gravitationally bound dust clump \citep{Visser+2021}. Altogether ices that end up in planetesimals could remain unaltered with respect to the nascent dust population during the final collapse stage of planetesimal formation. From this point of view, the effects of nonlocal ice processing implies a planetesimal population whose compositional gradients as a function of $r$ is smoother than what one would expect from the chemical gradients imposed by local condensation of ices. In particular, the chemical gradients of volatile ices around their respective ice lines could be shallower than expected from a sharp ice line. This could thus also be reflected in the initial volatile abundances of planetesimals.\\
Whether the volatile molecules comprising ices on dust are maintained inside the first generation of planetesimals crucially depends on the amount of radiogenic heating provided by $^{26}$Al. In the Solar System, devolatilization has only been averted in the outer Solar System, where planetesimal formation timescale exceeds the half-life of $^{26}$Al, ${\tau_{1/2}\approx 720}$ kyr \citep[e.g.,][]{Grimm&McSween1993, Monteux+2018}. However, some planetary systems may have lower $^{26}$Al abundance than the Solar System, which could result in significant retention of volatiles in planetesimals that formed early in the inner few AU of the planet-forming disk \citep{Lichtenberg+2019}.\\
Altogether it is clear that planetesimals can only maintain their original volatile composition in very specific cases. Furthermore it becomes clear that the only Solar System objects for which the volatile ice budget can be determined and that may have retained some similarity to the volatile abundances found in this work are comets. There is some evidence that suggests that comets may have formed through the collapse of gravitationally bound dust clumps \citep{Blum+2017, Lorek+2018, Nesvorny+2019}, and considerable evidence that ices in comets may have been inherited from interstellar ices to some degree \citep{Altwegg+2017, Rubin+2020, Bergner+2021}. However, direct comparison of the ice abundances predicted in this work with atmospheric cometary molecular abundances reported in e.g. \cite{Bockelee-Morvan+2017} or \cite{Rubin+2020} may still be misleading since the abundances of molecules in cometary atmospheres may not necessarily be representative for the ice composition of the cometary interior \citep[][]{Marboeuf+2014, Prialnik2014, Pajola+2017}.\\ 
Altogether, it is likely an oversimplification to directly map the ice composition predicted in our model onto the compositions of comets. However, the results shown in this paper could aid in estimates for primitive nucleus compositions used in cometary outgassing models \citep[e.g.,][]{Marboeuf+2014}.

%% file: TextFiles/5Conclusion.tex
\label{sec:5}

In this work, we have used a large number of monomer "tracer particles" calculated with the SHAMPOO code \citep{Oosterloo+2023} to quantitatively constrain the influence of transport processes on local volatile CHNOS budgets. We here binned the position timesteps associated with the trajectories of individual monomers on a preexisting background grid. This allowed for the construction of local populations of monomers that probe the properties of local dust affected by nonlocal disk processes. We summarize our key findings as follows:
\begin{itemize}
    \item In a massive disk where the dynamics of dust are dominated by diffusion throughout most of the disk, individual units of dust mass can be transported over significant distances. Spatially, diffusion results in local dust to be more nonlocal in the outer disk Fig. \ref{fig:31NonLocalF3_cleaned3OriginDiagram} and \ref{fig:31NonLocalF3_cleaned3OriginHists}. Specifically, at $r=50$ AU, individual monomers were found to originate from as far in as $r\approx 35$ AU and as far out as $r\approx 70$, while monomers at $r=2$ AU were found to originate from as far out as $r\approx 7$ AU. However, due to the closer spacing of ice lines at smaller $r$, we do see more mixing of dust from chemically different regions in the inner disk compared to the outer disk (Fig. \ref{fig:31NonLocalF3_cleaned3ChemicalOrigin}).     
    \item Ice processing via adsorption and photodesorption of individual monomers only occurs during $\lesssim 1-10$ \% of the entire evolutionary history, which means that most of the time, individual units of dust mass are located inside dust aggregates, insensitive to the local chemical composition of the gas phase (Fig. \ref{fig:32NonLocalF3_cleaned3Exposedness}).
    \item Nonlocal disk processing can have a significant effect on the absolute amounts of individual ices, in particular at smaller radial positions ($r=3,6$ AU) and for specific species (CO$_2$ and the sulfur-bearing species), while the relative ice mantle composition remains comparable to local disk processing (Fig. \ref{fig:33NonLocalF3_cleaned3NonlocalHistograms} and Fig. \ref{fig:33NonLocalF3_cleaned3NonLocalComposition}). The exact effect of nonlocal disk processing on the amount of ice of individual species depends on the position in the disk and chemical species.  
    \item Microscopic collisional mixing, the mixing of solid material throughout dust aggregates through cycles of coagulation and fragmention, can have a significant impact on the amount of ice associated with individual species near their respective ice line (Fig. \ref{fig:34MidplaneIcesComparison} and Fig. \ref{fig:34NonLocalF3_cleaned3AggregateIceDistributions}). This phenomenon occurs if the collisional timescale is comparable to or shorter than the adsorption or desorption timescale. In this work we derived that this can result in more than two orders of magnitude less ice with respect to a model where ice evolution remains limited to aggregate surfaces. Since diffusion of molecules out of the aggregate was assumed to be instantaneous in this work and diffusion effects are likely not negligible in the same disk region, this likely reflects a lower limit on the amount of ice that can be maintained in dust by this mechanism.    
\end{itemize}
Altogether the framework presented in this work provides valuable insights in the effects of nonlocal ice processing on the volatile CHNOS composition and distribution throughout dust aggregates. Suggested additional applications of the framework presented in this work include studying the effects of a time-evolving class I disk with accretional outbursts, and more evolved class II disks where significant dust settling and drift is expected to occur. Furthermore, coupling of our model results to planetesimal formation models could provide new insights in the initial ice budgets of plantesimals. Future modelling efforts are also recommended to explore regimes where molecule escape from aggregates is less efficient than currently considered, while another key area of exploration is the behavior of this process for different filling factors and fragmentation thresholds. 

%% file: TextFiles/6Appendix.tex
\onecolumn
\section{Chemical composition of the background disk model}
\label{sec:AA}
\input{TextFiles/SectionA/SectionAA.tex}
\newpage
\twocolumn
\section{Modifications to ice evolution treatment}
\label{sec:AB}
\input{TextFiles/SectionA/SectionAB.tex}

\section{Environment variation due to monomer discretization}
\label{sec:AC}
\input{TextFiles/SectionA/SectionAC.tex}

\section{Distribution of monomer data points}
\label{sec:AD}
\input{TextFiles/SectionA/SectionAD.tex}

\section{Dynamical behaviour throughout the sampling region}
\label{sec:AE}
\input{TextFiles/SectionA/SectionAE.tex}

%% file: TextFiles/SectionA/SectionAA.tex
\begin{figure}[ht!]
    \centering
    \includegraphics[width=.99\textwidth]{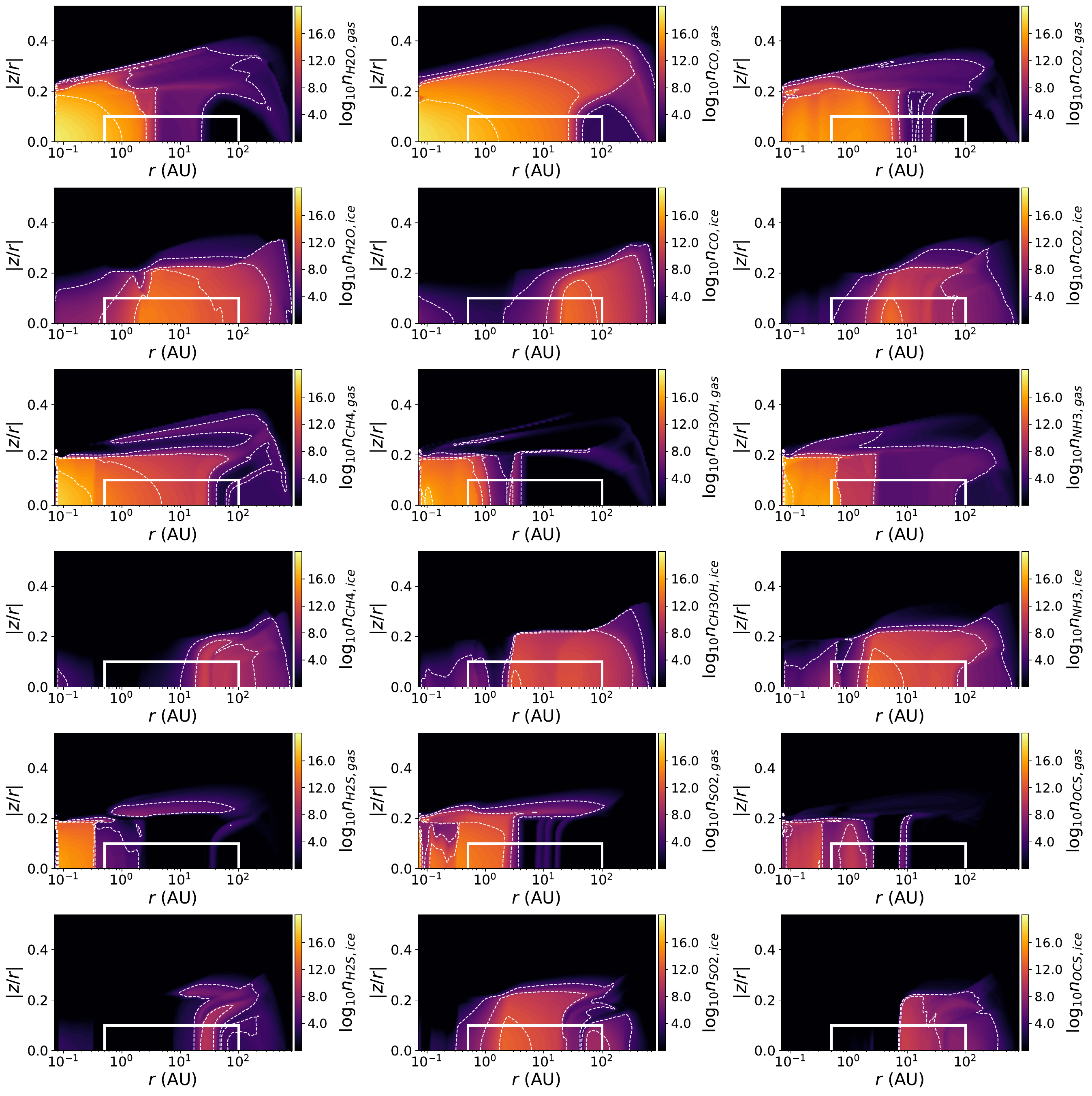}
    \caption{Number densities $n_x$ of gas and ice phase molecules in the background disk model. The number densities are presented in units of $\log_{10}(\text{molecules}\,\unit{m}{-3})$. The white boxes indicate the monomer sampling region defined in Sect. \ref{sec:2.3}, and the white dashed lines are abundance contours corresponding to the labels at the color bar.}
    \label{fig:AABackgroundModelAbundances}
\end{figure}

%% file: TextFiles/SectionA/SectionAB.tex
With respect to the version of SHAMPOO presented in \cite{Oosterloo+2023}, there have been significant improvements in the numerical stability of the ice evolution code, which originally displayed frequent convergence issues in the disk region between $r=1$ AU and $r=2$ AU. Furthermore, the treatment of thermal desorption in the version of SHAMPOO has been expanded upon. Specifically, the calculation of the thermal desorption rate now also includes first-order desorption \citep{Cuppen+2017}, where we use the same approach as used \cite{Woitke+2009}. In the extended prescription, the specific thermal desorption rate $\mathcal{R}_{\text{tds},x}$ for species $x$ is given by
\begin{align}
    \mathcal{R}_{\text{tds},x}=
       k_{\text{tds},x}N_{\text{des},x}m_x. 
\end{align}
Here, $m_x$ denotes the molecular mass of species $x$, and $N_{\text{des},x}$ denotes the number of molecules of species $x$ available for thermal desorption per unit of monomer area. $k_{\text{tds},x}$ denotes the molecule desorption rate \citep{Tielens&Allamandola1987, Cuppen+2017}
\begin{align}
    k_{\text{tds},x}=\nu_x\exp\left(-\frac{E_{\text{ads},x}}{k_\text{B}T_\text{d}}\right),
\end{align}
with $E_{\text{ads},x}$ denoting the adsorption energy of species $x$ and $T_\text{d}$ denoting the dust temperature. Furthermore, $\nu_0$ denotes the lattice vibration frequency given by \citep{Tielens&Allamandola1987}
\begin{align}
    \nu_0=\sqrt{\frac{2N_\text{ads}E_{\text{ads},x}}{\pi^2m_x}}.
\end{align}
Here, $N_\text{ads}$ denotes the number of molecular sites available for thermal desorption per ice monolayer per unit surface area. Furthermore, $N_{\text{des},x}$ depends on the desorption regime \citep{Woitke+2009}:
\begin{align}
\label{eq:newdesorption}
    N_{\text{des},x}=\begin{cases}
        N_\text{ML}N_\text{ads}f_x&\quad \text{if $N_\text{ice}\geq N_\text{act}$, 0th order desorption}\\
        \frac{M_x}{4\pi s_\text{m}^2 m_x}&\quad \text{if $N_\text{ice}<N_\text{act}$, 1st order desorption}.
    \end{cases}
\end{align}
Here, $N_\text{ML}$ denotes the number of ice monolayers available for desorption, $f_x=M_x/(m_xN_\text{ice})$ the fractional abundance of molecules of species $x$ in the monomer ice mantle. $M_x$ denotes the total mass of ice associated with species $x$ in the monomer ice mantle. $N_\text{ice}$ denotes the total number of ice molecules in the entire monomer ice mantle, calculated via
\begin{align}
    N_\text{ice}=\sum\limits_x^\text{all ices}\frac{M_x}{m_x}.
\end{align}
Furthermore, $N_\text{act}$ denotes the total number of sites available for thermal desorption on the monomer
\begin{align}
    N_\text{act}=4\pi s_\text{m}^2N_\text{ML}N_\text{ads}
\end{align}
In the first order desorption regime, the monomer ice mantle contains less molecules than the desorption sites available in the monomer ice mantle, which means that the thermal desorption rate is not limited by the number of available desorption sites, but the number of molecules on the monomer surface. 

%% file: TextFiles/SectionA/SectionAC.tex
\begin{figure*}[ht!]
    \centering
    \includegraphics[width=.99\textwidth]{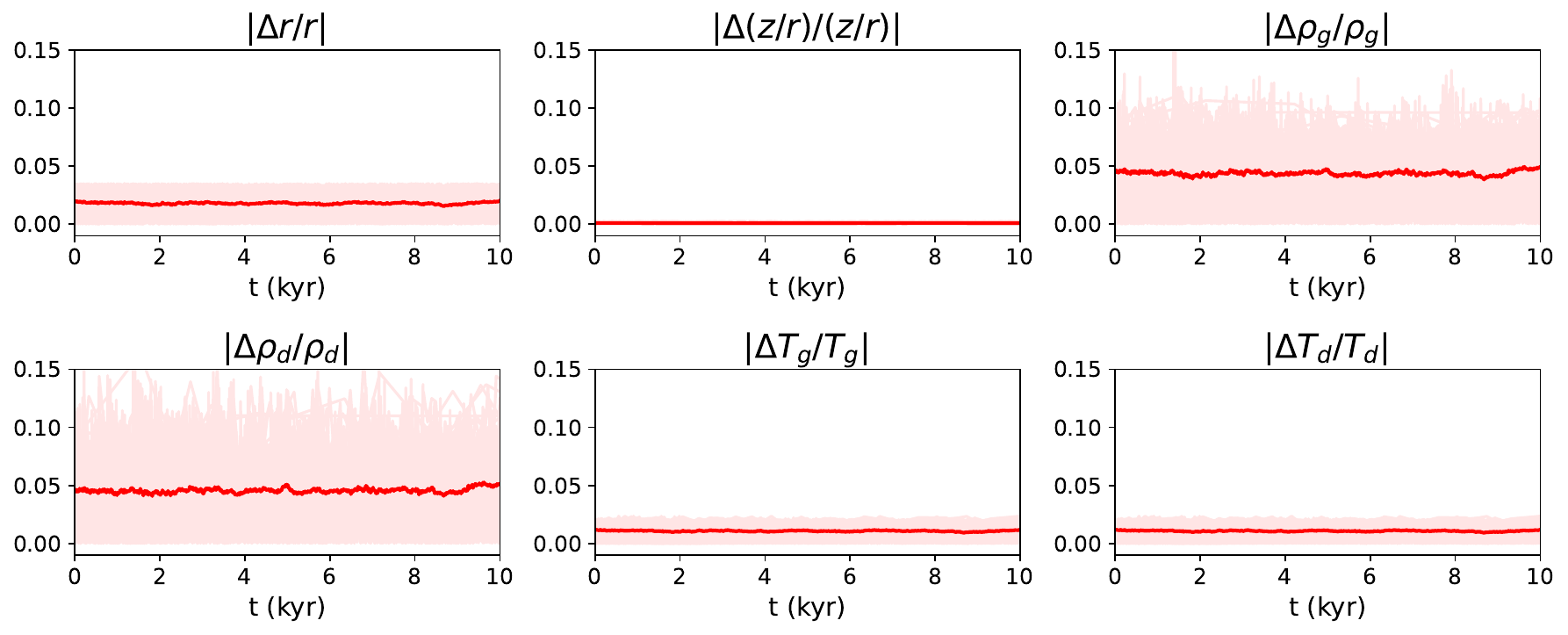}
    \caption{Relative difference for $r$, $\rho_\text{g}$, $\rho_\text{d}$, $T_\text{g}$, $T_\text{d}$ and absolute difference for $z$ (in AU) between 100 individual monomer trajectories and the value at the nearest background model grid point. The pink lines denote the trajectories of the individual monomers, while the red line denotes the average difference over all monomers.}
    \label{fig:ACDiscretizedCoordinatesError}
\end{figure*}
Monomers are usually not located exactly at the grid point coordinate $r_i,z_j$ used to define the spatial grid. In particular for larger $r$, the spatial grid cells associated with the default ProDiMo grid can become multiple AU in size, which raises the question how representative the interpolated physical disk conditions at the monomer position $r,z$ are for the physical conditions at grid point $r_i,z_j$. Since we utilize the monomer as a representative particle $r_i,z_j$, this difference in physical conditions induces an uncertainty in how representative the monomer in reality is for the physical conditions at $r_i,z_j$. Fig. \ref{fig:ACDiscretizedCoordinatesError} presents the relative difference between the interpolated physical conditions at the positions of 100 monomers randomly chosen from the nonlocal monomer set defined above and the physical conditions associated with the grid cell $r_i,z_j$ in which the monomers have been binned at every time step. The upper left and upper center panel denote the relative difference between the original monomer coordinates $r,z$ and the grid point coordinates $r_i,z_j$ of the binning grid cell associated with the monomer. It is clear that the relative difference between $r$ and $r_i$ is approximately $2\%$, while closer inspection of the upper center panel reveals that vertically the difference in $z/r$ and $z_j/r_i$ is $\sim 0.1\%$. However, it becomes clear from the upper right panel and the lower left panel of Fig. \ref{fig:ACDiscretizedCoordinatesError} that there are significant errors in the gas and dust densities, respectively, with monomers being located at a position where $\rho_\text{g}$ and $\rho_\text{d}$ differ on average by $\sim 5\%$ from the densities in the associated grid cells, with outliers up to $\sim 15\%$. $\rho_\text{g}$ and $\rho_\text{d}$ crucially affect the dynamical and collisional evolution through the inverse linear dependence of the Stokes number on $\rho_\text{g}$ and dependence of the collision rates on $\rho_\text{d}$. The spread in temperature appears to be smaller, with the average monomer associated with a grid cell usually having an associated gas and dust temperature that deviates $\sim 2\%$ on average from the grid cell temperature. The gas temperature primarily affects collision rates for monomers in small aggregates through Brownian motion, whereas it also affects the adsorption rate for exposed monomers, which means that its effects on the monomer population as a whole, of which a large fraction resides unexposed in large dust aggregate interiors are likely limited. However, the dust temperature affects the thermal desorption rate which depends exponentially on the inverse of $T_\text{d}$, which means that a relative variation in $T_\text{d}$ would cause the thermal desorption rate to vary by a factor 2.6. Altogether we can thus expect considerable spread in monomer ice composition due to the binning of the position coordinates. This means that the ice composition inferred from the monomer set for a specific grid point in the background model reflects the average ice composition for the range of physical conditions possible in the entire grid cell, which does not neccessarily have to be identical to the average ice composition under the chemical conditions at point $r_i,z_j$ in the background disk model. The effects of the coordinate discretization on the dynamical and collisional history appear to be more limited due to their comparatively weaker dependence on the background model physical conditions.

%% file: TextFiles/SectionA/SectionAD.tex
\begin{figure*}[ht!]
    \centering
    \includegraphics[width=\textwidth]{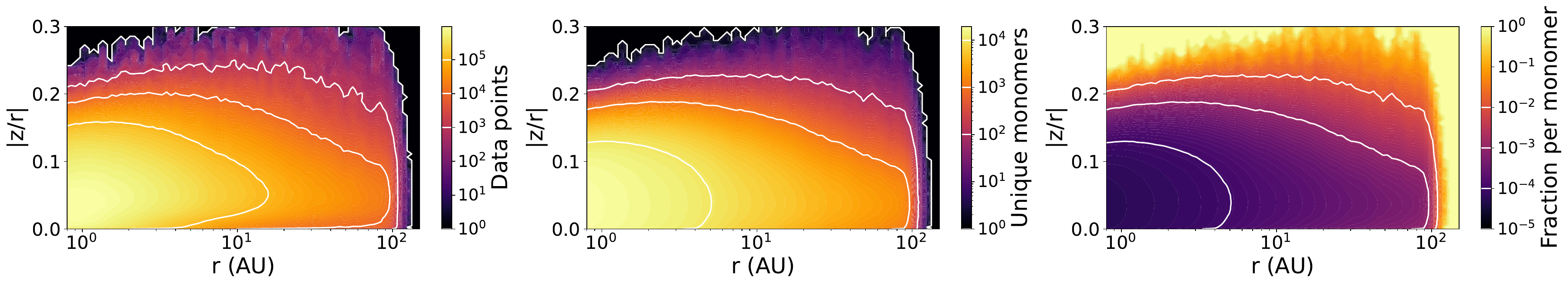}
    \caption{Distribution of the monomer population over the background model grid. Left panel: Spatial distribution of the number of timesteps (data points) per spatial grid cell, based on a simulation of 64000 monomers. Center panel: Number of unique monomers that have visited a given cell during their evolutionary trajectories. Right panel: The average fraction of grid cell data points contributed by a single unique monomer.}
    \label{fig:ADNonLocalF3_cleaned3DataPoints}
\end{figure*}
In this section we explore the distribution of the data points associated with the monomers in the nonlocal simulation defined in Sect. \ref{sec:2.3} over the background model grid. We here aim to spatially limit the disk regions where the nonlocal monomer set defined in Sect. \ref{sec:2.3} populates grid cells with a statistically significant number of data points. Fig. \ref{fig:ADNonLocalF3_cleaned3DataPoints} presents the spatial distribution of the number of data points per grid cell, the number of unique monomers that has visited the concerning grid cell, and the average number of data points per monomer in every particular grid cell. \\
It becomes clear from Fig. \ref{fig:ADNonLocalF3_cleaned3DataPoints} that for all grid cells in our region of analysis, $r\in [1,100]$ AU, $z/r\in [-0.1,0.1]$, the number of data points is larger than $10^3$, usually exceeding even $10^4$ for most grid cells. It is also key to investigate the number of unique monomers associated with the data points, since individual monomers can contribute more than one data point to grid cells (see also Fig. \ref{fig:23MonomerStatisticsExample}). This could in principle allow a large fraction of the data points in a grid cell to be associated with the trajectory of a single unique monomer. However, Fig. \ref{fig:ADNonLocalF3_cleaned3DataPoints} shows that except for close to 100 AU, the data in grid cells originates from the trajectories of $10^3$ to $10^4$ unique monomers. The resulting fraction of data points per grid cell originating from a unique monomer is usually below $10^{-3}$, while in the inner disk between $r=1$ AU and $r=4$ AU to $6$ AU near the midplane up to $|z/r|=0.1$, there is a region where the fractional contribution of a single monomer trajectory to the full data set of grid cells is below $10^{-4}$. Anticipating the results of Sect. \ref{sec:3.1} and Sect. \ref{sec:3.4}, the inner disk will likely display larger variations in ice composition, and hence a larger data set and monomer population size will be required to obtain reliable expectation values for ice amounts. It is also clear that throughout the entire sampling region, individual monomers usually contribute only a small fraction to the full data set of a grid point. Therefore it is unlikely that any of the data sets in grid cells will be biased towards the trajectory of a single monomer dominating the data set.

%% file: TextFiles/SectionA/SectionAE.tex
\begin{figure}[ht!]
    \centering
    \includegraphics[width=.45\textwidth]{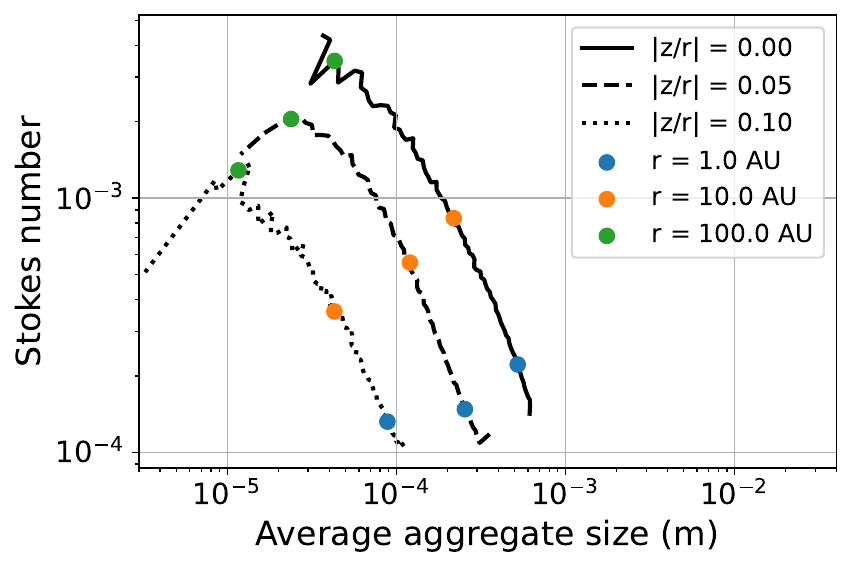}
    \caption{Radial behavior of the Stokes number associated with $\langle s_\text{a} \rangle$ throughout the disk for different values of $|z/r|$.}
    \label{fig:AENonLocalF3_cleaned3StokesNumbers}
\end{figure}

The dynamical behavior of monomers is crucially affected by the Stokes numbers of their home aggregate. Specifically, the Stokes number associated with the expected home aggregate size does provide information on whether the dynamical behavior of dust at a given position in the disk is typically governed by aerodynamic drag or by turbulent diffusion. We therefore explore the Stokes number associated with the expected value for $s_\text{a}$ at a number of values of $|z/r|$ throughout the monomer sampling region in Fig. \ref{fig:AENonLocalF3_cleaned3StokesNumbers}. It becomes clear that the Stokes number increases as a function of $r$ for any of the values of $|z/r|$ considered. This might be counter-intuitive at first glance since the average aggregate size $s_\text{a}$ decreases as a function of $r$. However, we note that the increase in the Stokes number is primarily a consequence of the strongly decreasing gas density and to a lesser extent also the decreasing sound speed \citep{Birnstiel+2012, Oosterloo+2023}. We find the Stokes number attains its highest value, $\text{St}\approx 3\e{-3}$, at the outer boundary of our sampling region in the midplane at $r=100$ AU. Altogether this means that although settling and drift may have a small effect on the dynamical transport of monomers located in the largest aggregates at larger radial distances, the motion of aggregates is primarily affected by particle diffusion mediated by gas turbulence. We note that this behavior is a consequence of the young, massive disk assumed in the background disk model. This gives rise to higher gas densities and hence lower Stokes numbers compared to a disk where settling and drift dominate the dynamical behavior of aggregates.
